\documentclass{raa}            
\usepackage{multirow}
\usepackage{graphicx,times}             
\usepackage{natbib}
\usepackage{amssymb,amsmath}
\bibpunct{(}{)}{;}{a}{}{,}

\begin{document}

  \title{Pulsar glitch activities: the spin parameters approach}

   \volnopage{Vol.0 (20xx) No.0, 000--000}      
   \setcounter{page}{1}          

   \author{Innocent Okwudili Eya 
      \inst{1,2,3}
      \and Evaristus Uzochukwu Iyida
      \inst{2,3}   
   }

   \institute{ Physics/Electronics Technique -- Department of Science Laboratory Technology, University of Nigeria, Nsukka, Nigeria. {\it innocent.eya@nn.edu.ng}\\
        \and
             Department of Physics and Astronomy, University of Nigeria, Nsukka, Nigeria.\\
        \and
             Astronomy and Astrophysics Research lab, University of Nigeria, Nsukka, Nigeria.\\            
\vs\no
   {\small Received 20xx month day; accepted 20xx month day}}

\abstract{Glitch activity refers to the mean increase in pulsar spin frequency per year due to rotational glitches. It is an important tool for studying super-nuclear matter using neutron star interiors as templates. Glitch events are typically observed in the spin frequency ($\nu$) and frequency derivative ($\dot{\nu}$) of pulsars. The rate of glitch recurrence decreases as the pulsar ages, and the activity parameter is usually measured by linear regression of cumulative glitches over a given period. This method is effective for pulsars with multiple regular glitch events. However, due to the scarcity of glitch events and the difficulty of monitoring all known pulsars, only a few have multiple records of glitch events. This limits the use of the activity parameter in studying neutron star interiors with multiple pulsars. In this study, we examined the relationship between the activity parameters and pulsar spin parameters (spin frequency, frequency derivative, and pulsar characteristic age). We found that a quadratic function provides a better fit for the relationship between activity parameters and spin parameters than the commonly used linear functions. Using this information, we were able to estimate the activity parameters of other pulsars that do not have records of glitches. Our analysis shows that the relationship between the estimated activity parameters and pulsar spin parameters is consistent with that of the observed activity parameters in the ensemble of pulsars. 
}

   \authorrunning{I. O. Eya \& E. U. Iyida }            
   \titlerunning{Pulsar glitch activities and the spin parameters}  

   \maketitle

%
%
\section{Introduction}           
\label{sect:intro}
 Pulsar glitches are sudden changes in the rotation rate of a spinning neutron stars that emit beams of electromagnetic radiation \citep{Reichley1969,Radhakrishnan1969,Chukwude2010,yu2013,Lai2016,lower2021,B2022,Grover2023}. These glitches have gained significant interest in astrophysics as they provide a means of probing the interior of neutron stars \citep[e.g.][]{Lyne1992,b53,Andersson2012,eya17a,Haskell2018,eya2019}. This is because the physics underlying the phenomenon, which is connected to the behaviour of matter in neutron star interior, helps to understand the behaviour of matter in exotic states beyond that obtainable in the terrestrial environments.
The mean change in pulsar spin frequency per year due to glitches known as glitch activity parameter (and hereafter, the activity parameter) is normally quantified by the regression of cumulative glitch sizes in a given pulsar with respect to the time those glitches were observed \citep{mckenna90}. 
This approach is readily feasible and reliable in pulsars with multiple regular glitches. 
However, the paucity of glitch events in many pulsars is now a bane to a comprehensive understanding of the nature of neutron star interior via glitch activity parameters.
Hence, there is a need to devise other means of quantifying the expected activity parameters of pulsars \citep[see][for a recent attempt]{eya22a}. 
If that is done, it will give an insight into the expected activity parameter of pulsars without glitches. 

Though hundreds of glitches have been observed across the population of neutron stars \citep[see, for example][]{Chukwude2010,b7,yu2013,lower2021,B2022,Li2023} the number of pulsars with records of glitch is still very small compared to numbers of known pulsars\footnote{One can see the ATNf pulsar catalogue for number of known pulsar https://www.atnf.csiro.au/research/pulsar/psrcat.}.
Large glitches $\Delta\nu/\nu \gtrsim 10^{-7}$ are often characterized by spectacular step changes in spin frequency $(\nu)$, usually accompanied by a change in frequency derivative $(\dot{\nu})$ \citep{Antonelli2023a}.
It is known that glitches are not due to sudden changes in the external electromagnetic torques on the neutron star \citep{b29,b3,Haskell2015}. 
Therefore, glitch models are built on the neutron star crust quake mechanism or the neutron star interior superfluid vortex distribution and angular momentum transfer within the neutron star \citep{b29,b3,b2,Zhou14,eya17a,eya20}. 
There has been no lack of theories to explain pulsar glitch trigger mechanisms and subsequent post-glitch behaviour \citep[see, for example,][and for reviews \citep{Haskell2015,eya2020,zhouetal2022,Antonopoulou2022}]{b4,b5,alpar95,Xiao2011,Zhou14,Lai2016,eya20,rencoret2021}.
Despite that, there has not been a consensus on the actual origin of glitches in neutron stars. 

\begin{figure}
\centering
	\includegraphics[scale=0.8]{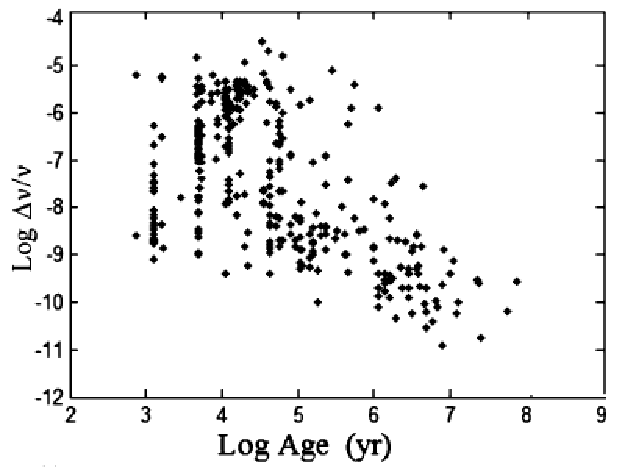}
    \caption{The distribution of glitch sizes with age. }
    \label{fig:age_size_figure}
\end{figure}

Many analysts have focused on the statistical study of glitch events either in individual pulsars or in ensembles of pulsars \citep[eg.][]{melatos08,lyne00,b7,eya14,Fuentes2017,eya2019a,eya22a}.
One of the major findings of this approach that have stood the test of time and is about
 reaching a status of standalone is that the number, frequency, and sizes of glitches decrease as the pulsar age \citep[][and as also shown in Figure ~\ref{fig:age_size_figure}]{eya17b,eya22a,B2022}.
In that, it could be said that the trigger of glitches in pulsars is age-dependent.  
This notion favours the glitch mechanism that hinges on neutron star interior superfluid vortex dynamics and transfer of angular momentum\citep[e.g.][]{Alpar1981,b2,eya17a}. 
Pulsars do not rotate as a rigid body. Instead, the crust rotates at a different speed than the interior superfluid, which creates a rotation lag between the two \citep{b3,Alpar1981,b2,eya17a}. 
This lag allows angular momentum to be stored in the superfluid, carried by vortices. As the pulsar spins down, the magnitude of the rotation lag increases. Eventually, the vortices can no longer withstand the stored momentum, causing them to unpin, migrate, and transfer angular momentum to the crust, resulting in a spin-up phenomenon known as a glitch.
After a glitch, the vortices gradual repin. 
The repining characterizes the relaxation of pulsars after glitch known as glitch recovery \citep{yu2013}. 
Young pulsars with higher spin-down rates are more susceptible to glitches due to the larger rotation lag. Additionally, the interiors of younger pulsars are believed to be hotter, and vortex dynamics are highly temperature-dependent \citep{Alpar1989}. 
Thus, due to the turbulent conditions found in hot environments, combined with the high rate at which young pulsars slow down, vortex pinning and unpinning occur more frequently in younger pulsars. This can result in more instances of glitch events compared to older pulsars.
As the pulsar ages and its spin-down rate decreases, the likelihood of glitches also decreases \citep{Haskell2015,eya17a}.
The pulsar characteristic age ($\tau_{c} = \frac{\nu}{-2\dot\nu}$), which is somewhat a quotient of the pulsar spin frequency and the frequency derivative is the spin-down time of a given pulsar from birth believing that pulsars behave like a dipole rotator in a vacuum. 
The frequency and the frequency derivative are the primary parameters altered in glitches. 
Characterizing the alterations in them is a major way of studying glitch events in pulsars.
As we proceed, in Section (\ref{sec:Distribution of glitches}) we shall see how glitches are also distributed in the plane of the spin frequency and its derivative. 
This will help in visualizing the spin parameter that will be reliable enough in estimating the pulsar's glitch activities. 

Another vital product of statistical analysis of glitch events is the ``activity parameter" \citep{mckenna90}.
In a statistical study of 48 glitches from 18 pulsars, \citet*{lyne00} found that the activity parameter is directly proportional to the pulsar frequency derivative. 
\citet{wang00}, from an analysis of 76 glitches from 25 pulsars found that there is no consistent relationship between glitch magnitude and the time since the previous glitch or the time to the following glitch, either for the ensemble or for individual pulsars. 
As such, drivers of glitch activity must be external to the pulsars, thereby being in line with \citet*{lyne00} report.
This finding of \citet{wang00} is buttressed by that of \citet{eya2019a} who showed that there is no significant difference between inter-glitch time interval proceeding large glitches and that of small glitches. 
In that, the drivers of glitches are tied to a parameter, which is time-dependent and the plausible parameter is the $ \dot{\nu} $.
In the earliest study that estimated activity parameters of pulsars, \citet{urama99} studied 71 glitches from 30 pulsars. They showed that for middle-age pulsars ($\tau_c \sim 10^4 - 10^5\  \rm{yr}$), the glitch activity parameter is proportional to the logarithm of the frequency derivative, $|\dot{\nu}|$. 
Based on the proportionality, they were able to estimate the glitch activity, $a_g$, for all the sampled youthful pulsars as:
$   a_g \approx 41.4 + 3.22\ \rm{Log}\,|\dot{\nu}|.$
The glitches studied in these cases above have magnitudes $(\Delta\nu/\nu \sim 10^{-6} - 10^{-9})$  and represent $ \sim $ 10\% of glitches reported to date.
 As many more glitches are being detected in the conventional radio pulsars, and other manifestations of neutron stars and as glitch size range is wider now ($ \Delta\nu/\nu \sim 10^{-5} - 10^{-11} $). 
  There is, therefore, the need to re-examine the relationship between the glitch activity and the pulsars parameters altered in glitches.

Meanwhile, \cite{Fuentes2017} in a study of 348 glitches in the rotation of 141 neutron stars, using the approach of integrated glitch activity\footnote{That is collective glitch activity for collections of pulsars sharing a common property.} and laying emphasis on the absolute glitch spin-up size ($\Delta \nu$) noted that the glitch activity correlates with spin frequency $ \nu $, and it's derivative $|\dot{\nu}| $.
Also, they noted that the activity parameter also correlates with parameters, which are a combination of $\nu $ and $\dot{\nu} $ such as characteristic age and spin-down luminosity, but not with magnetic field. 
In their analysis, pulsars were grouped with respect to their spin parameters and estimated their glitch activity. 
As such, the result is based on averaging over objects of similar spin properties.

In this analysis, we shall focus on individual pulsars in quantifying the activity parameter. 
We are to work with both the absolute ($ \Delta\nu $) and fractional glitch sizes ($ \Delta\nu /\nu$).
The activity parameter --- spin parameter relationship shall be explored to obtain a suitable equation for estimating the activity parameters of a given pulsar.
The estimated activity parameter shall be tested for consistency with the already-known relationship between the activity parameter and pulsar spin parameters.

\section{Data}
The pulsar spin parameters are from \cite{manchester2005} and updated with the Australian Telescope National Facility pulsar catalogue (ATNF)\footnote{https://www.atnf.csiro.au/research/pulsar/psrcat \label{atnf}} and references therein. 
We selected all the pulsars from the ATNF pulsar catalogue that possess both $ \nu $ and $ \dot{\nu} $ measurements, that are not recycled/millisecond pulsars, not in a binary system or accreting matter from its environment. 
This results in a total of 2215 pulsars and 21 Magnetar/AXPs. However, it is important to note that our analysis is focused on pulsars and not Magnetar/AXPs.

The glitch parameters are from \cite{b7,B2022} and updated with the Jodrell Bank Observatory pulsar glitches\footnote{http://www.jb.man.ac.uk/pulsar/glitches/original.html, and http://www.jb.
man.ac.uk/pulsar/glitches.html.} and references therein. 
For a large compilation of $ \Delta \nu $ one can readily access http://www.jb.man.ac.uk/pulsar/glitches/original.html.
The 670 glitches reported as of the time of this analysis came from 208 neutron stars, representing $ \sim 7\ \rm{\%} $ of the neutron star population (see footnote~\ref{atnf}). 
About a third of these glitches came from just eight pulsars: PSRs J0534$+$2200 (Crab Pulsar) -- 30 glitches, J0537$-$6910 -- 53 glitches, J0631$+$1036 --17 glitches, J0835$-$4510 (Vela Pulsar) -- 24 glitches, J1341$-$6220 --35 glitches, J1413-6141 -- 14 glitches, J1740$-$3015 -- 36 glitches, and J1801$+$2304 -- 15 glitches. 
Most of the glitching pulsars (nearly two-thirds of them) have a record of just one glitch each. 
This few number of glitches per pulsar seriously limits the number of pulsars available for statistical analysis. 
For a sample to ascertain the relationship between the activity parameters and the spin parameters, we selected pulsars that have undergone at least five (5) glitches.
There are a total of 31 pulsars in this group.  
These 31 pulsars account for half of the total number of observed glitches. 
Their characteristic ages, $\tau_c$, are in the 
range $1.26 \times 10^3$ to $\sim 1.38 \times 10^6 $ \rm{yr}.

\section{Distribution of glitches across pulsar's spin parameters altered in glitches}
\label{sec:Distribution of glitches}
\begin{figure}
\centering
	\includegraphics[scale = 0.8]
	{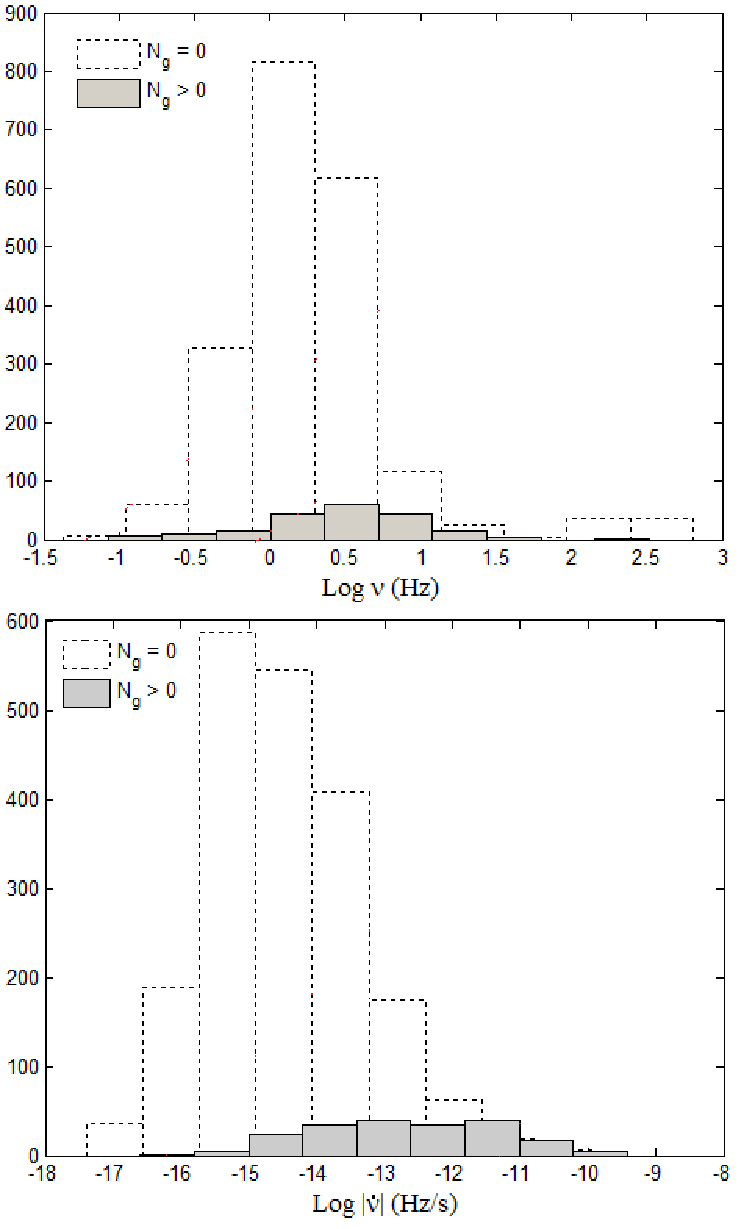}
    \caption{Top panel: distribution of pulsars spin frequency. Bottom panel distribution of pulsars frequency derivative. 
    $ N_{g} $ denote number of glitches.}
    \label{distribution_spin_paramter_glitch2}
\end{figure}

It is worth noting that glitches in pulsars are seen in their spin frequency and frequency derivative. 
These parameters are those that are primarily altered in glitches. 
In that, processes culminating in glitches could be connected to them. 
To investigate whether the presence or absence of a glitch in a given pulsar is biased by the magnitude of its spin frequency or the derivative, distributions of the pulsars spin frequency and that of the spin-down rate are examined concerning pulsars with glitches and those without glitches. 
This is shown in Figure ~\ref{distribution_spin_paramter_glitch2}. 
The shaded histograms are for objects with records of glitches, whereas the unshaded are for objects without\footnote{Note: many of these pulsars might have glitched but were not observed} a record of glitch event.

Upon examining the top panel (horizontally), it can be observed that there is no significant difference in the spin frequency magnitude between pulsars that have experienced glitch events and those that have not. 
The range of spin frequencies for pulsars with glitches is merely a subset of the range of spin frequencies for pulsars without glitches.
The majority of pulsars with glitches have their spin frequency in the range of Log $ \nu $ (1 --10) Hz and centred at $ \sim $ Log $ \nu $ = 0.5, while there is also a large concentration of pulsars without glitches in that range. 
This is a pointer that if a glitch event is solely dependent on pulsars' spin frequency, pulsars of this spin frequency range are potential glitch candidates. 
As such, one can extrapolate findings obtained in pulsars with glitches to those without glitches. 
In the bottom panel (Figure~\ref{distribution_spin_paramter_glitch2}), it appears that the distribution of pulsars with glitches is skewed with respect to the distribution of those without glitches. 
The proportion of pulsars with glitches to those without glitches concerning the bin size decreases with decreasing frequency derivative. 
This show that pulsar of high frequency derivative are more prone to glitches than the lesser one. 
If one is to compare this with that of the spin frequency, it is a pointer that the trigger of glitches could be external to the pulsars as the frequency derivative is a function of external torque on the pulsars.     
As such, pulsars with high-frequency derivatives are likely to glitch compared to pulsars with lower-frequency derivatives.
This is in line with most glitch mechanisms, which hinge on the spinning down of pulsars. 
Pulsar spin-down is a prime contributor to processes that culminate in glitches.

\begin{figure}
\centering
	\includegraphics[scale=0.9]{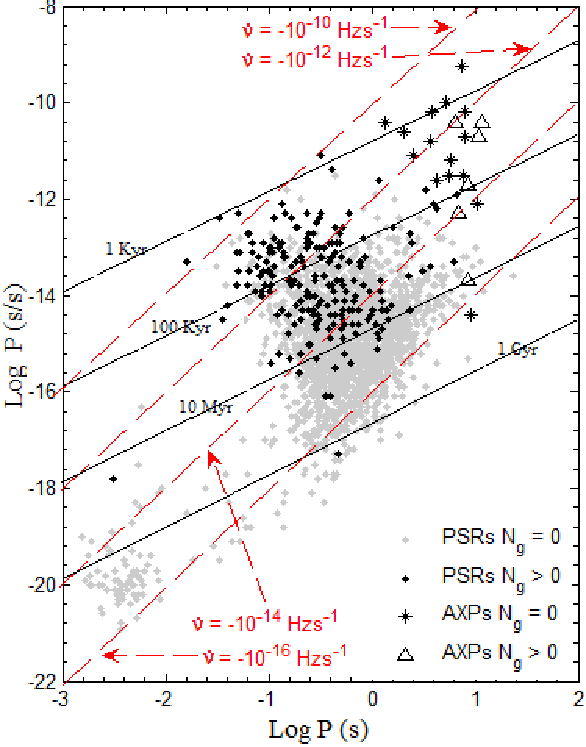}
    \caption{The Period -- Period derivative diagram. The black solid line is a line of constant characteristic age (spin-down time). The dashed red lines are lines of constant frequency derivative. The horizontal axis is $ \nu^{-1} $}
    \label{p_p diagram}
\end{figure}

To investigate how pulsars with glitches and those without glitches are distributed simultaneously in the plane of spin frequency ($ \nu = P^{-1} $) and its derivatives, we explore the period (P) - Period derivative ($ \dot{P} $) diagram. 
The $ P$ -- $\dot{P} $ diagram is a powerful tool to examine the evolution of pulsars at a glance. 
The diagram is shown in Figure ~\ref{p_p diagram}.  
From it, at a glance, it is seen that recorded glitch events are mostly from pulsars of $ \tau_{c} < 10 \,\rm{Myr}$  and $ \dot{\nu} > -10^{-14}\, \rm{Hzs^{-1}}$.
All pulsars of $ \dot{\nu} \geq $ $ -10^{-10}\ \rm{Hzs^{-1}}$ have recorded a glitch. 
Though there are very few pulsars in this category.
Equally, if one moves down perpendicularly across the lines of constant $ \dot{\nu} $, the ratio of pulsars with glitches to those without glitches decreases.  
This observation is another indication that glitch events are tied to the magnitude of the pulsar frequency derivative. 
That is younger pulsars with high-frequency derivatives have shown more glitch events than older pulsars with low-frequency derivatives.

\section{The Glitch activity parameter} 
The glitch activity parameter describes the rate and size of glitches in a given pulsar.
It measures the frequency and size of glitches that occur in pulsar rotation for a given period. 
It is often defined as the mean fraction increase in rotation frequency per year due to a glitch or number of glitches per unit time \citep{mckenna90,urama99}. 
In pulsar with multiple glitches, the $ i^{th}$ spin-up sequence is readily characterised with ($t_{i} $,  $\Delta\nu_{i} $) or ($t_{i} $, $ \frac{\Delta\nu_{i}}{\nu} $) where  $t_{i} $ is the inter-glitch time interval of $ i^{th}$ glitch, while $\Delta\nu_{i} $ and  $ \frac{\Delta\nu_{i}}{\nu} $ are the absolute and fractional glitch size for $t_{i} $ glitch respectively.
Following \citet{Dibetal2008}, the absolute glitch activity parameter, $A_g$, is defined as:
\begin{equation}
A_g = \frac{1}{\Delta t_g}\sum^{N_{g}}_{i=1}{\Delta\nu_{i}}.
\label{eq:ABS}
\end{equation}
where $\Delta t_g$ is the total time over which the glitches were observed and the sum is over all
glitches; while the fractional glitch activity, $a_g$, is
\begin{equation}
a_g = \frac{1}{\Delta t_g}\sum^{N_{g}}_{i =1}{\frac{\Delta\nu_{i}}{\nu}},
\label{eq:FGS}
\end{equation}
In these definitions, it is assumed that the statistical properties of the sequence ($t_{i}, \Delta \nu_{i}$) remain constant regardless of the observation window. 
However, this may not always be the case when the observation time is unknown (or irregular observation), and only the sequence ($t_{i}, \Delta \nu_{i}$) is available. 
To address this and in order not to over estimate the activity parameters, a qualitative modifications has been made \citep{Montoli2021,Antonelli2023b} resulting in
\begin{equation}
A_g \approx \frac{N_{g} - 1}{N_{g}(t_{N_{g}} - t_{1})} \sum^{N_{g}}_{i=1}{\Delta\nu_{i}},
\label{eq:ABS2}
\end{equation}
and as such,
\begin{equation}
a_g \approx \frac{N_{g} - 1}{N_{g}(t_{N_{g}} - t_{1})} \sum^{N_{g}}_{i =1}{\frac{\Delta\nu_{i}}{\nu}}.
\label{eq:FGS2}
\end{equation}
$ N_{g} $ is the number of glitches, $ t $ is the time of the glitch.

In this analysis, we adopted the use of Equations ~\ref{eq:ABS2} and \ref{eq:FGS2} in estimating the glitch activity parameters.
The spin parameters and the activity parameters of the sampled pulsars are presented in Table~(\ref{tab:activity_table}).
As in Table~(\ref{tab:activity_table}) and as also seen in other analyses \citep[for example see][]{urama99,lyne00,b7,eya22a}, the magnitude of the activity parameter in a given pulsar is mainly dominated by the number of large glitches ($ \Delta\nu/\nu > 10^{-7}$) in such a pulsar at that period. 
This is because the addition of let's say glitch sizes of $ \Delta\nu/\nu > 10^{-9}$ to $ \Delta\nu/\nu > 10^{-7}$ will require up to hundreds of such small glitches to effectively alter the magnitude of $ \Delta\nu/\nu > 10^{-7}$.
As such, the magnitude of glitch activity to an extent is just a measure of the preponderance of large glitches in a given pulsar. 
In addition, we have to note that the magnitude of activity parameters in individual pulsars has not been a constant value.
It varies from analysis to analysis, though revolves around a narrow range with minimal dispersion in pulsars with regular and similarly sized glitches. 
The variations are mainly due to an unequal observational time span and an unequal number of large glitches involved.

\begin{table*}
	\small
	\centering
	\caption{Table showing the spin parameters and the observed activity parameter of the sampled pulsars.}
	\label{tab:activity_table}
	\begin{tabular}{lccccccc} 
		\hline
		Pulsar &$ \nu $& $ \vert\dot{\nu}\vert $ $  $ &$ \tau_{c} $  &$ N_{g} $ &$ N_{L} $  &a$_{g} $  & A$_{g} $\\
		J name &(Hz) & ($ 10^{-12} $ Hz $\rm{s^{-1}} $) &(kyr)&	& &($ 10^{-7} \rm{yr^{-1}} $)&($ 10^{-7} \rm{Hz\,yr^{-1}} $)\\
		\hline
0205+6449	&	15.22	&	44.87	&	5.37	&	9	&	7	&	7.18	&	109.30	\\
0534+2200	&	29.95	&	377.54	&	1.26	&	30	&	2	&	0.21	&	6.40	\\
0537-6910	&	62.03	&	199.23	&	4.93	&	53	&	46	&	6.57	&	407.64	\\
0631+1036	&	3.47	&	1.26	&	43.6	&	17	&	2	&	2.20	&	7.63	\\
0729-1448	&	3.97	&	1.79	&	35.2	&	6	&	1	&	3.29	&	13.05	\\
0742-2822	&	6.00	&	0.60	&	157	&	9	&	1	&	1.17	&	7.03	\\
0835-4510	&	11.19	&	15.67	&	11.3	&	24	&	20	&	7.32	&	81.89	\\
1016-5857	&	9.31	&	7.01	&	21	&	5	&	4	&	6.21	&	57.79	\\
1023-5746	&	8.97	&	30.88	&	4.6	&	7	&	6	&	14.84	&	133.11	\\
1048-5832	&	8.08	&	6.28	&	20.4	&	6	&	3	&	4.77	&	38.54	\\
1105-6107	&	15.82	&	3.97	&	63.2	&	5	&	3	&	1.34	&	21.17	\\
1341-6220	&	5.17	&	6.77	&	12.1	&	35	&	20	&	7.14	&	36.94	\\
1357-6429	&	6.02	&	13.05	&	7.31	&	5	&	0	&	6.30	&	37.96	\\
1413-6141	&	3.50	&	4.09	&	13.6	&	14	&	10	&	4.84	&	16.94	\\
1420-6048	&	14.67	&	17.89	&	13	&	7	&	7	&	5.52	&	80.89	\\
1617-5055	&	14.42	&	28.09	&	8.13	&	6	&	3	&	1.03	&	14.79	\\
1709-4429	&	9.76	&	8.86	&	17.5	&	5	&	5	&	2.77	&	26.99	\\
1731-4744	&	1.21	&	0.24	&	80.4	&	6	&	3	&	1.11	&	1.34	\\
1740-3015	&	1.65	&	1.27	&	20.6	&	36	&	11	&	3.09	&	5.09	\\
1801-2304	&	2.40	&	0.65	&	58.3	&	15	&	6	&	2.38	&	5.72	\\
1801-2451	&	8.00	&	8.20	&	15.5	&	7	&	6	&	5.09	&	40.74	\\
1803-2137	&	7.48	&	7.52	&	15.8	&	6	&	4	&	5.95	&	44.53	\\
1814-1744	&	0.25	&	0.05	&	84.6	&	7	&	0	&	0.07	&	0.02	\\
1825-0935	&	1.30	&	0.09	&	233	&	7	&	1	&	0.06	&	0.08	\\
1826-1334	&	9.85	&	7.31	&	21.4	&	7	&	5	&	3.64	&	35.90	\\
1841-0524	&	2.24	&	1.18	&	30.2	&	7	&	4	&	1.44	&	3.23	\\
1902+0615	&	1.48	&	0.02	&	1380	&	6	&	0	&	0.00	&	0.00	\\
1952+3252	&	25.30	&	3.74	&	107	&	6	&	0	&	1.36	&	34.36	\\
2021+3651	&	9.64	&	8.89	&	17.2	&	5	&	5	&	4.75	&	45.79	\\
2225+6535	&	1.47	&	0.02	&	1120	&	5	&	1	&	0.45	&	0.65	\\
2229+6114	&	19.37	&	29.37	&	10.5	&	9	&	7	&	2.69	&	52.15	\\									

\hline
\end{tabular}\\
Note: $ N_{g} $ denotes number of observed glitches, $ N_{L} $ denotes number of large glitches ($ \Delta\nu/\nu \gtrsim 10^{-7}$). 
\end{table*}

\begin{figure}
\centering
	\includegraphics[scale=0.8]{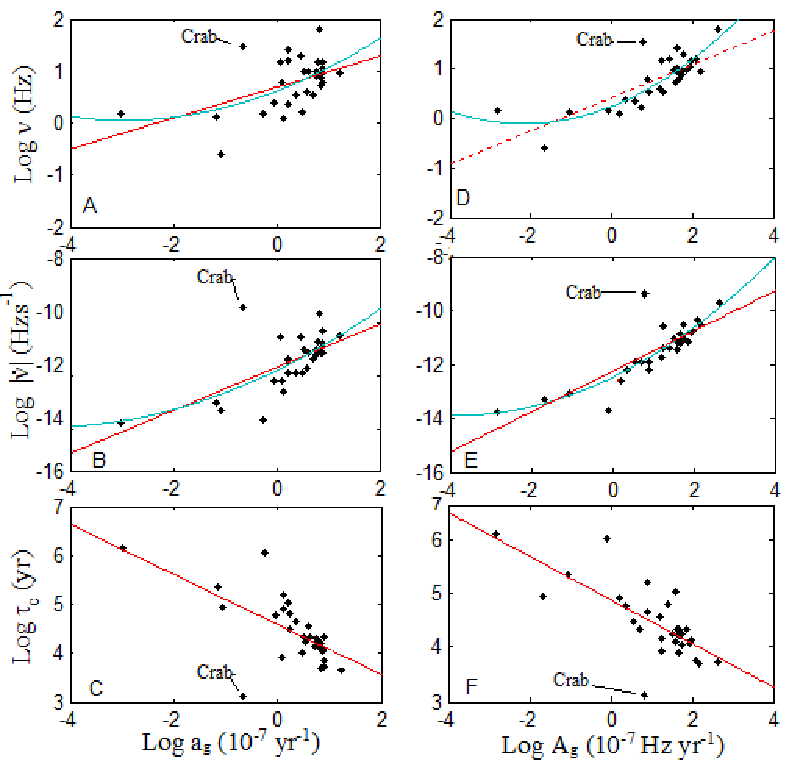}
    \caption{Plots of the pulsar spin parameters, $\nu$, $|\dot\nu|$ and $\tau_c$ with both the fractional glitch activity ($a_g$) and the absolute glitch activity ($A_g$).}
    \label{fig:ag-spindown_figure}
\end{figure}

\begin{table}
    \centering
    \caption{Summary of the statistical parameters of the fits in Figure \ref{fig:ag-spindown_figure} parameters}
\label{tab:fit_parameters}
    \begin{tabular}{|l|l|c|c|c|}
        \hline
        Figure&Fits&r&$ R^{2} $&$n_{res} $\\
        \hline
        \multirow{2}{*}{\ref{fig:ag-spindown_figure}A $_{(a_{g}-\nu)}$} & quadratic &0.58&0.34&3.732 \\
        & linear &0.50&0.25&3.974 \\
        \hline
        \multirow{2}{*}{\ref{fig:ag-spindown_figure}B $_{(a_{g}-|\dot{\nu}|)}$} & quadratic &0.85&0.72& 2.440\\
        & linear &0.67&0.42&3.484 \\
        \hline
        {\ref{fig:ag-spindown_figure}C $_{(a_{g}-\tau_{c})}$} 
        & linear & -0.65 &0.43&3.468 \\
        \hline
        \multirow{2}{*}{\ref{fig:ag-spindown_figure}D $_{(A_{g}-\nu)}$} & quadratic &0.81&0.66& 3.732\\
        & linear &0.78&0.61&3.974\\
        \hline
        \multirow{2}{*}{\ref{fig:ag-spindown_figure}E $_{(A_{g}-|\dot{\nu}|)}$} & quadratic &0.91&0.83& 2.163\\
        & linear &0.83&0.70&3.503 \\
        \hline
        \ref{fig:ag-spindown_figure}F $_{(A_{g}-\tau_{c})}$
        & linear &-0.73 &0.54& 4.351 \\
        \hline
    \end{tabular}
    \\
  $n_{res}$ denotes norm of residuals
\end{table}

As the glitch events alter\footnote{The alteration could be permanent or transient} the pulsar's spin frequency and its derivatives, and the event rate depends on the pulsar characteristic age, these parameters are likely tools in studying the activity parameters.
In what follows, we probe the relationship between the activity parameters and these pulsar spin parameters.
To do this, we start by plotting the parameters.
The plots are shown in Fig.~\ref{fig:ag-spindown_figure}. 
From the pattern of the data points in the logarithm plots [see Figs.~\ref{fig:ag-spindown_figure} (A, B, D and E)], two physical functions can be used to describe the relationships: one is linear and the other is of a polynomial of second order (quadratic). 
The relationship with characteristic age is just linear [as seen in Figs.~\ref{fig:ag-spindown_figure} (C and F)].
The summary of the statistical parameters of the fits are presented in Table \ref{tab:fit_parameters}. 
The correlation coefficient is denoted by the symbol $r$, while the coefficient of determination is denoted by $R^{2}$. 
The norm of residuals is denoted by $n_{res}$. 

The coefficient of determination, $R^{2}$, is a statistical measure that indicates the percentage of the variance in the dependent variable that can be explained by the independent variables.  
In simpler terms, it shows how well a model fits a given data. 
A higher $ R^2 $ indicates a better fit.
In this context, it shall determine the percentage of the activity parameter that can be determined from the pulsar spin parameters. 
However, it is always recommended to use this metric in combination with other diagnostic tools and domain knowledge to have an informed understanding of the model's performance and reliability. 
Therefore, it is important to also examine the norm of residuals, to determine the quality of the fits. 

The norm of residuals is a statistical measure that calculates the differences between the observed values and the values predicted by a statistical model. 
It represents the differences between the actual data points and the corresponding values predicted by the model. 
In order to select the best model that fits the data, the norm of residuals is used as a criterion for comparison between different models. 
A lower norm of residuals typically suggests a better fit.
By analyzing both $R^{2}$ and $n_{res}$, we can assess the goodness of fit of the models and make better decisions.
 
 In Figure~\ref{fig:ag-spindown_figure} (A), the correlation coefficients (see Table~\ref{tab:fit_parameters}) suggest a moderate correlation between the $ a_g $ and $ \nu $.  
An increase in $\nu$ is positively associated with increases in $a_{g}$.
However,  the magnitude of $ R^{2}$ in both fits ($ < 0.5 $) indicates that the pulsar spin frequency could not account for up to 50\% variance in the $ a_g $ that is explained by the pulsar spin frequency. 
 Nonetheless, the two possible relationships envisaged from Figure \ref{fig:ag-spindown_figure} (A) are described by equations of the forms:
\begin{equation}
\rm{Log}\, a_ {_g} \simeq -0.539 \,(\rm{Log}\,\nu)^2 + 1.663\, (\rm{Log}\, \nu) - 0.296
\label{equ_FGS_F0_quad}
\end{equation}
for the quadratic nature, while the linear nature takes the form.
\begin{equation}
\rm{Log}\, a_ {_g} \simeq 0.989 \,(\rm{Log}\, \nu) - 0.251.
\label{equ_FGS_F0_linear}
\end{equation}
In Figure ~\ref{fig:ag-spindown_figure} (B) the $ r = 0.85 $ indicates a very strong relationship between the $ a_g $, and $ |\dot{\nu}| $ in a quadratic model and a strong relationship ($ r = 0.67 $) in the linear model. 
The relationships are described by equations of the forms:
\begin{equation}
\rm{Log}\, a_ {_g} \simeq -0.2067 \,(\rm{Log}\, |\dot{\nu}|)^2 -4.239\, (\rm{Log}\, |\dot{\nu}|) - 20.79,
\label{equ_FGS_F1_quad}
\end{equation}
from the quadratic nature, and the linear nature is
\begin{equation}
\rm{Log}\, a_ {_g} \simeq 0.5196 \,(\rm{Log}\, |\dot{\nu}|) + 6.2453.
\label{equ_FGS_F1_linear}
\end{equation}
The value of $R^{2}$, which is 0.72 in the quadratic nature, indicates that about 70 \% of $a_{g}$ relies on the pulsar spin frequency derivative. 
In contrast, the linear nature has an $R^{2}$ value of 0.42, which means that only about 40 \% of the variance of $a_{g}$ can be explained by $|\dot{\nu}|$ in the linear model. 
This suggests that the model's deviation from the observed data could go up to 60 \%. 
Moreover, the norm of residuals shows a difference of an order of magnitude between the two models (refer to Table  \ref{tab:fit_parameters}) with that of the quadratic being the lesser one. 
This demonstrates that the quadratic nature model fits the observed $a_{g}$ better than the linear nature model.
In Figure ~\ref{fig:ag-spindown_figure} (C), the relationship has the form:
\begin{equation}
\rm{Log}\, a_ {_g} \simeq -0.8298 \,(\rm{Log}\, \tau_c) + 3.9710.
\label{equ_FGS_character_age_linear}
\end{equation}
 The value of $ a_{_g} $ is inversely proportional to $ \tau_c $, with a correlation coefficient of $ r = -0.65 $. 
 This indicates a strong inverse relationship between the two variables, the coefficient of determination $ R^{2} $ is equal to 0.42.
 This means that only around 40 \% of the variability in $ a_{_g} $ can be explained by $ \tau_{c} $, suggesting that around 60 \% information on the activity parameter could be lost if one relies solely on pulsar characteristic age in estimating the glitch activity parameter. 
In intriguing the inverse relationship between $ a_{_g} $ and $ \tau_{c} $ and the consistent of this result with the linear nature of Figure ~\ref{fig:ag-spindown_figure} (B), it's worth noting that $ \tau_c \approx -\nu/2\dot{\nu} $. 
The inverse\footnote{consequence of $ \dot{\nu} $ being at denominator} relationship and the consistency indicate that the effects of $ |\dot{\nu}| $ on the activity parameter are more pronounced than those of $ \nu $. 
Therefore, $ |\dot{\nu}| $ is a better indicator of the occurrence of glitch events in pulsars.

On the other hand, the absolute glitch activity -- pulsar spin parameter relationship appears to be tighter than that fractional glitch activity relationship (as seen in Table \ref{tab:fit_parameters}). 
With the pulsar spin frequency, the relationships are as follows:
In Figure \ref{fig:ag-spindown_figure} (D)
\begin{equation}
\rm{Log}\, A_ {_g} \simeq -0.539 \,(\rm{Log}\,{\nu})^2 +2.663\, (\rm{Log}\, \nu) - 0.296,
\label{equ_ABS_F0_quad}
\end{equation}
is for the quadratic nature, while  
\begin{equation}
\rm{Log}\, A_ {_g} \simeq 1.989\,(\rm{Log}\, \nu) - 0.251
\label{equ_ABS_F0_linear}
\end{equation}
is for the linear nature.
Unlike the Figure \ref{fig:ag-spindown_figure} (A), $ A_g $ exhibit a strong relationship with $ \nu $. 
In both natures, $ \nu $ could account for up to 60 \% of $ A_{g} $ as demonstrated by the magnitude of $ R^{2} $ (see Table \ref{tab:fit_parameters}).
In Figure \ref{fig:ag-spindown_figure} (E), the relationship is of the form
\begin{equation}
\rm{Log}\, A_ {_g} \simeq -0.187 \,(\rm{Log}\,|\dot{\nu}|)^2 -3.360\, (\rm{Log}\, |\dot{\nu}|) - 12.554
\label{equ_ABS_F1_quad}
\end{equation}
for the quadratic nature, while the linear nature is
\begin{equation}
\rm{Log}\, A_ {_g} \simeq 1.0939 \,(\rm{Log}\, |\dot{\nu}|) + 13.683.
\label{equ_ABS_F1_linear}
\end{equation}The magnitude of $r$ and $ R^{2} $ indicate a very strong relationship between the parameters (see Table \ref{tab:fit_parameters}). 
This is consistent with the findings in Figure \ref{fig:ag-spindown_figure} (B) but with an even stronger correlation. 
This emphasizes the significance of $ |\dot{\nu}| $ in determining the glitch activity of pulsars. 
It shows that in modelling the glitch activity of pulsars without records of glitch events, spin frequency derivative should be the first and foremost parameter to consider. 
Moreover, the quadratic nature of the relationship with $R^{2} = 0.83$ and the least $ n_{res} $ suggest that the best-fitted model between the activity parameter and the spin parameters is the quadratic model.
This finding we shall examine further as we proceed.
Finally, the $ A_g $ relationship with $ \tau_c $ (Figure \ref{fig:ag-spindown_figure} F) is linear and inversely.
It has an equation of the form:
\begin{equation}
\rm{Log}\, A_ {_g} \simeq -1.3011 \,(\rm{Log}\, \tau_c) + 6.8376
\label{equ_ABS_charac_age_linear}
\end{equation}

Meanwhile, it is worth noting that the data point of the Crab pulsar deviated significantly from the trends defined by data points of other pulsars. 
On isolating the Crab pulsar and evaluating the correlation coefficients, the magnitude rosed by 5 \%. 
This observation we suggest to be a consequence of the preponderance of small-sized glitches in Crab pulsar.
This could indicate that how the activity parameter of Crab pulsar relates to its spin parameters is different from others or that the glitch mechanism in Crab pulsar is different from others.

\section{Estimating the activity parameters of pulsars with multiple glitches}
Using Equations (\ref{equ_FGS_F0_quad}), (\ref{equ_FGS_F0_linear}),  (\ref{equ_ABS_F0_quad}) and  (\ref{equ_ABS_F0_linear})  
the activity parameters of the sampled pulsars are calculated from the pulsar spin frequency.
The values are presented in Table (\ref{tab:ag_predict_freq}).
Then using Equations (\ref{equ_FGS_F1_quad}), (\ref{equ_FGS_F1_linear}), (\ref{equ_ABS_F1_quad},) and  (\ref{equ_ABS_F1_linear}),  
the activity parameters are calculated from the pulsar spin frequency derivatives, while Equations (\ref{equ_FGS_character_age_linear}) and (\ref{equ_ABS_charac_age_linear}) are used to calculate the activity parameters from the pulsar characteristic age. 
The values are presented in Table (\ref{predict_freq_deri}) and (\ref{tab:ag_predict_age}) respectively.

\begin{table}
\small
	\centering
	\small
	\caption{Table showing the activity parameter estimated from the pulsar spin frequency ($\nu$). Columns for quadratic nature are headed with (quad) while that of linear nature is headed with (linear).
	Note: $ A_g $ is in unit of $10^{-7}Hz\,  yr^{-1}$  while $ a_{g} $ is in unit of $ 10^{-7} $ $yr^{-1}$.}
	\label{tab:ag_predict_freq}
	\begin{tabular}{lcccc} 
		\hline
	Pulsar	&$ a_{g} $	&$ a_{g} $	&	$ A_{g} $	&	$ A_{g}$	\\
	
J name	&	(quad)	&	(linear)	&	(quad)	&	(linear)\\	
\hline
0205+6449	&	8.26	&	8.29	&	125.62	&	126.08	\\
0534+2200	&	9.65	&	16.18	&	288.88	&	484.69	\\
0537-6910	&	8.98	&	33.25	&	556.73	&	2062.68	\\
0631+1036	&	2.79	&	1.92	&	9.70	&	6.68	\\
0729-1448	&	3.21	&	2.20	&	12.77	&	8.73	\\
0742-2822	&	4.69	&	3.30	&	28.14	&	19.78	\\
0835-4510	&	7.17	&	6.12	&	80.24	&	68.47	\\
1016-5857	&	6.45	&	5.10	&	60.04	&	47.47	\\
1023-5746	&	6.30	&	4.91	&	56.50	&	44.07	\\
1048-5832	&	5.88	&	4.43	&	47.52	&	35.82	\\
1105-6107	&	8.38	&	8.61	&	132.57	&	136.25	\\
1341-6220	&	4.13	&	2.85	&	21.38	&	14.74	\\
1357-6429	&	4.71	&	3.31	&	28.35	&	19.94	\\
1413-6141	&	2.81	&	1.94	&	9.85	&	6.78	\\
1420-6048	&	8.14	&	7.99	&	119.33	&	117.18	\\
1617-5055	&	8.08	&	7.86	&	116.48	&	113.26	\\
1709-4429	&	6.63	&	5.34	&	64.75	&	52.12	\\
1731-4744	&	0.68	&	0.67	&	0.82	&	0.81	\\
1740-3015	&	1.09	&	0.92	&	1.80	&	1.51	\\
1801-2304	&	1.82	&	1.34	&	4.37	&	3.21	\\
1801-2451	&	5.84	&	4.39	&	46.75	&	35.14	\\
1803-2137	&	5.57	&	4.11	&	41.66	&	30.71	\\
1814-1744	&	0.03	&	0.14	&	0.01	&	0.04	\\
1825-0935	&	0.77	&	0.73	&	1.00	&	0.95	\\
1826-1334	&	6.67	&	5.39	&	65.74	&	53.12	\\
1841-0524	&	1.66	&	1.25	&	3.73	&	2.80	\\
1902+0615	&	0.94	&	0.83	&	1.40	&	1.23	\\
1952+3252	&	9.47	&	13.70	&	239.49	&	346.49	\\
2021+3651	&	6.58	&	5.28	&	63.47	&	50.85	\\
2225+6535	&	0.92	&	0.82	&	1.35	&	1.20	\\
2229+6114	&	8.94	&	10.52	&	173.26	&	203.77	\\
		\hline
	\end{tabular} 
\end{table}

\begin{table}
	\centering
	\small
	\caption{Table showing the activity parameter estimated from the pulsar spin frequency derivative ($|\dot{\nu}|$). Columns for quadratic nature are headed with (quad) while that of linear nature is headed with (linear).
	Note: $ A_g $ is in unit of $10^{-7} \rm{Hz\,  yr^{-1}}$  while $ a_{g} $ is in unit of $ 10^{-7} $ $\rm{yr^{-1}}$.}
	\label{predict_freq_deri}
	\begin{tabular}{lcccc} 
		\hline
	Pulsar	&$ a_{g} $	&$ a_{g} $	&	$ A_{g} $	&		$ A_{g}$	\\
	
J name	&	(quad)	&	(linear)	&	(quad)	&	(linear)\\	
\hline
0205+6449	&	8.74	&	7.38	&	155.26	&	230.24	\\
0534+2200	&	6.32	&	22.32	&	318.42	&	2366.96	\\
0537-6910	&	7.59	&	16.02	&	277.33	&	1176.22	\\
0631+1036	&	2.42	&	1.16	&	8.93	&	4.64	\\
0729-1448	&	3.04	&	1.38	&	12.91	&	6.78	\\
0742-2822	&	1.40	&	0.79	&	3.83	&	2.07	\\
0835-4510	&	7.60	&	4.27	&	82.96	&	72.83	\\
1016-5857	&	5.97	&	2.81	&	45.52	&	30.21	\\
1023-5746	&	8.51	&	6.08	&	126.88	&	153.02	\\
1048-5832	&	5.72	&	2.66	&	41.59	&	26.79	\\
1105-6107	&	4.69	&	2.09	&	27.93	&	16.21	\\
1341-6220	&	5.89	&	2.76	&	44.25	&	29.09	\\
1357-6429	&	7.26	&	3.89	&	73.07	&	59.65	\\
1413-6141	&	4.76	&	2.13	&	28.69	&	16.75	\\
1420-6048	&	7.82	&	4.58	&	90.69	&	84.21	\\
1617-5055	&	8.41	&	5.79	&	120.11	&	137.95	\\
1709-4429	&	6.48	&	3.18	&	54.80	&	39.03	\\
1731-4744	&	0.61	&	0.48	&	1.15	&	0.75	\\
1740-3015	&	2.43	&	1.16	&	8.94	&	4.64	\\
1801-2304	&	1.49	&	0.82	&	4.20	&	2.25	\\
1801-2451	&	6.31	&	3.05	&	51.58	&	35.85	\\
1803-2137	&	6.12	&	2.92	&	48.17	&	32.63	\\
1814-1744	&	0.10	&	0.21	&	0.10	&	0.13	\\
1825-0935	&	0.21	&	0.29	&	0.28	&	0.25	\\
1826-1334	&	6.06	&	2.87	&	47.07	&	31.61	\\
1841-0524	&	2.31	&	1.11	&	8.25	&	4.29	\\
1902+0615	&	0.02	&	0.12	&	0.02	&	0.04	\\
1952+3252	&	4.56	&	2.03	&	26.48	&	15.20	\\
2021+3651	&	6.49	&	3.18	&	54.97	&	39.20	\\
2225+6535	&	0.03	&	0.14	&	0.03	&	0.05	\\
2229+6114	&	8.46	&	5.92	&	123.27	&	144.84	\\

		\hline
	\end{tabular} 
\end{table}

\begin{table}
	\centering
	\small
	\caption{Table showing the activity parameter estimated from the pulsar characteristic age ($\tau_c$). 
	Note: Only linear fit is considered. $ A_g $ is in unit of $10^{-7} \rm{Hz\,  yr^{-1}}$  while $ a_{g} $ is in unit of $ 10^{-7} $ $\rm{yr^{-1}}$}.
	\label{tab:ag_predict_age}
	\begin{tabular}{lcc} 
		\hline
	Pulsar	&$ a_{g} $	&$ A_{g} $\\
	
J name	&	(linear)	&	(linear)\\	
\hline
0205+6449	&	7.51	&	149.81	\\
0534+2200	&	25.02	&	917.35	\\
0537-6910	&	8.07	&	166.70	\\
0631+1036	&	1.32	&	10.93	\\
0729-1448	&	1.58	&	14.28	\\
0742-2822	&	0.46	&	2.20	\\
0835-4510	&	4.05	&	59.11	\\
1016-5857	&	2.42	&	27.24	\\
1023-5746	&	8.54	&	181.78	\\
1048-5832	&	2.48	&	28.25	\\
1105-6107	&	0.97	&	6.87	\\
1341-6220	&	3.83	&	54.26	\\
1357-6429	&	5.82	&	101.88	\\
1413-6141	&	3.48	&	46.89	\\
1420-6048	&	3.61	&	49.61	\\
1617-5055	&	5.33	&	89.20	\\
1709-4429	&	2.82	&	34.21	\\
1731-4744	&	0.80	&	5.09	\\
1740-3015	&	2.46	&	27.90	\\
1801-2304	&	1.04	&	7.60	\\
1801-2451	&	3.12	&	39.82	\\
1803-2137	&	3.07	&	38.88	\\
1814-1744	&	0.76	&	4.77	\\
1825-0935	&	0.33	&	1.35	\\
1826-1334	&	2.39	&	26.61	\\
1841-0524	&	1.79	&	17.30	\\
1902+0615	&	0.08	&	0.15	\\
1952+3252	&	0.63	&	3.56	\\
2021+3651	&	2.86	&	34.96	\\
2225+6535	&	0.09	&	0.19	\\
2229+6114	&	4.31	&	64.79	\\
		\hline
	\end{tabular} 
\end{table}

\section{Comparing the estimated activity parameters with the observed}
A monotonic comparison of the estimated activity parameters of the sampled pulsars with the observed is summarised in Table \ref{tab:mono_test}. 
In all the comparison, a positive direct association is observed. 
\begin{table}
	\centering 
	\small
	\caption{Summary of correlation coefficient (r) between the estimated activity parameters and the observed.  $q$ and $l$ subscripts denote results from quadratic and linear nature respectively}
	\label{tab:mono_test}
	\begin{tabular}{lcccccc} 
		\hline
	&	$ a_{g} $	& \vline	&$ A_{g} $&		\\
	\hline
Parameter	used &	r	&\vline	&r	\\
\hline
$\nu_ {_q}$	&	0.57		&	\vline &0.81	\\
$\nu_{_l}$	&	0.50		&	\vline &0.73	\\
$|\dot{\nu}|_{_q}$	& 0.80	&\vline	&0.90	\\
$|\dot{\nu}|_{_l}$	& 0.67 &\vline	&0.84	\\
$\tau_{c}$	&	0.50	&\vline	&0.73	\\	
\hline
	\end{tabular} 
	
\end{table}
The highest correlation came from  $|\dot{\nu}|_q $, though in all the comparisons, the correlation coefficient are significant at 5 \% significant level.
Next, we examine the similarity in data point distributions using two-dimensional Kolmogorov -- Simnove (K--S) test and the kernel density estimator plot.
The K--S test helps to determine whether the distributions of the two quantities in question are similar and of the same continuous distribution, while the kernel density plot will present a picture of the similarity in the distributions measure of central tendency. 
The kernel density approach has an advantage over the use of histograms in the sense that the final look of the plot does not depend on the choice of the class interval like the histogram \citep{heumann2016} and the issue of lack of continuity that is sometimes posed by histograms is avoided.
In addition, the kernel density estimator plot provides a direct comparison between the width and peaks\footnote{modal and mean values} of the distributions under investigation.

In the K--S test, a null hypothesis is true (i.e. h = 0) is that the two distributions in question are similar. 
Now, with regards to the distributions of the estimated and observed glitch activity, if the null hypothesis is true, it indicates that the estimated glitch activity is in high precision with the observed glitch activity.
The alternative hypothesis (i.e. the null hypothesis is false. h = 1) is that the two distributions are significantly different.
As such, the estimated glitch activity is not similar to the observed glitch activities.
The P-value determines the reliability of the result (i.e. to ascertain if the null or alternative hypothesis is obtained by chance), and the K-value is just the maximum distance between the curves\footnote{the values of both P and K ranges from 0 to 1}. 
A large P-value is a key point to accept the null hypothesis, else otherwise. 
The test is done at a 5 \% significant level. 
The result is rejected outrightly for any $ P \leq 0.05 $.

\begin{figure}
\centering
\includegraphics[scale=0.8]{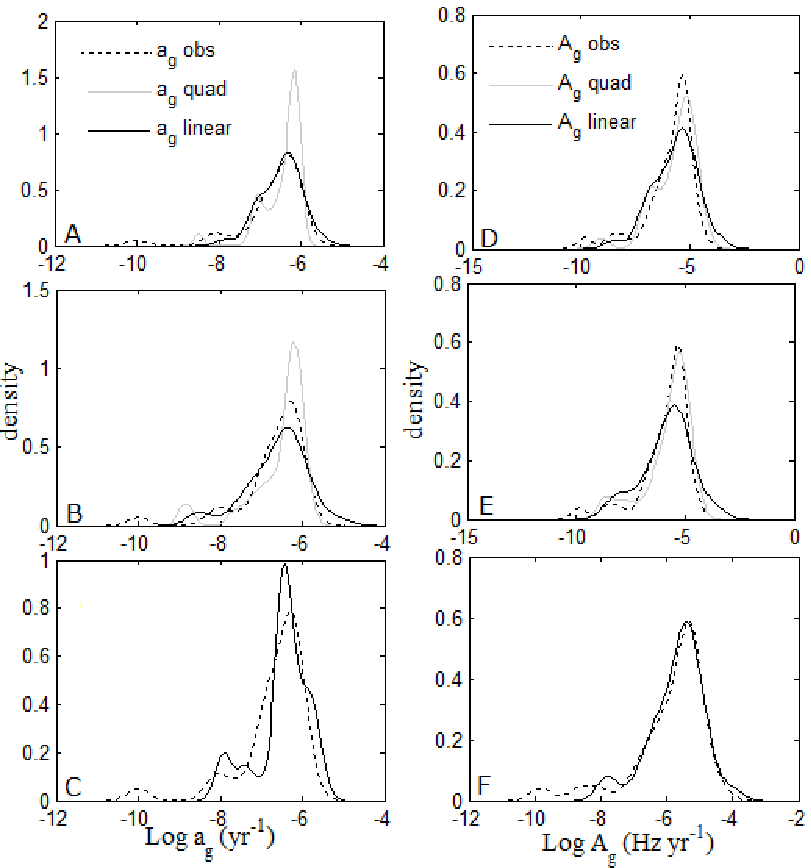}
    \caption{Kernel density of the distribution of observed and estimated glitch activities. Broken lines are for the observed values. The solid lines are for the estimated values: grey lines denote the quadratic nature, while black denotes in linear nature.}
    \label{fig:kernel_density}
\end{figure}

The summary of the K-S test results is presented in Table ~\ref{tab:KS_test}, while the kernel density plot is shown in Fig ~\ref{fig:kernel_density}. 
In the results, the null hypothesis is true in all (i.e. $ h = 0 $). 
This is an indication that the distributions of the observed and estimated activity parameters are similar (despite that the data point of the Crab pulsar foists an outlier), thereby supporting the result presented in Table (\ref{tab:mono_test}).
As such, each of the three spin parameters (i.e. $\nu, |\dot{\nu}|$ and $\tau_{c}$) could be used in estimating the glitch activity of pulsars. 
The magnitudes of the P-values also support that the null hypothesis is true is not by chance\footnote{the probability that the result is obtained by chance is (1 - P)}.
When we juxtapose the result from the K--S test with the result in Table (\ref{tab:mono_test}) and the magnitudes of the norm of residuals, we state that in estimating the activity parameters of pulsars, the best parameter to use is $ |\dot{\nu}| $ and it is best described by Equations (\ref{equ_FGS_F1_quad}) and (\ref{equ_ABS_F1_quad}) --- the quadratic nature\footnote{note that parameters are in logarithm.}.

Moreover, the Kernel density plots (Fig \ref{fig:kernel_density}) also support the outcome of the K--S test that the null hypothesis is true in all.
The widths are similar, and the peaks of the distributions are similarly valued.
The difference in densities is just a consequence of an unequal number of events with the same value. 
This similarity in shape indicates that the compared distributions have similar measures of central tendencies.
Then, the behavior of each of the distributions in isolation can be extrapolated to the others, and a similar result obtained.
From the $ A_g $ panel, the mean peak is -5.5, which corresponds to absolute glitch activity of $ \sim $ 63.1$ \times 10^{-7} \rm{Hz yr^{-1}} $, while from the $ a_g $ panel, the mean peak is -6.277 corresponding to fractional glitch activity of $ \sim $ 5.28 $ \times 10^{-7} \rm{yr^{-1}} $. 
These values represent the value where the activity parameters are centred respectively.

Meanwhile, for emphasis, a look at panels (E) and (F) sees the highest levels of agreement between the estimated activity parameter with the observed. 
In (E), it is between the quadratic nature and the observed, while in (F), it is between the linear nature and the observed.  
Knowing that $ \tau_c $ is a function of $ \dot{\nu} $, one can readily say that the observation in F is just an extrapolation from E. 
As such, the prime driver of these agreements is $ \dot{\nu} $ supporting the earlier choice made with the results of the K--S test concerning Equation (\ref{equ_ABS_F1_quad}).
\begin{table}
	\centering
	\small
	\caption{Table showing the K--S test results between the observed and estimated glitch activities from the spin parameters. h = 0 in all. q and l subscripts denote results from quadratic and linear nature respectively}
	\label{tab:KS_test}
	\begin{tabular}{lcccccc} 
		\hline
	&	$ a_{g} $	&	&\vline	&$ A_{g} $&		&		\\
	\hline
Spin parameter	&	P	&	K	&\vline	&P	&	K	\\
\hline
$\nu_ {_q}$	&	0.560	&0.194		&	\vline &0.778	&	0.161	\\
$\nu_{_l}$	&	0.778	&0.161		&	\vline &0.559	&	0.193	\\
$|\dot{\nu}|_{_q}$	&0.666	&0.193		&\vline	&0.954	&	0.129	\\
$|\dot{\nu}|_{_l}$	&0.778	&0.161	&\vline	&0.778	&	0.161	\\
$\tau_{c}$	&	0.363	&	0.226	&\vline	&0.998	&	0.097	\\	
\hline
	\end{tabular} 
	
\end{table}

\section{Activity parameters of pulsars without records of glitches}
Now we have established that the quadratic nature of the activity parameters with respect to pulsar spin frequency derivative are good enough to calculate the activity parameters of pulsars with multiple glitches, we proceed to estimate the activity parameters of pulsars without a record of glitches. 
The activity parameters of the 2215 pulsars selected from the ATNF pulsar catalogue are calculated. 
They are presented in the Appendix (Table A).

\begin{figure}
\centering
\includegraphics[scale=1.2]{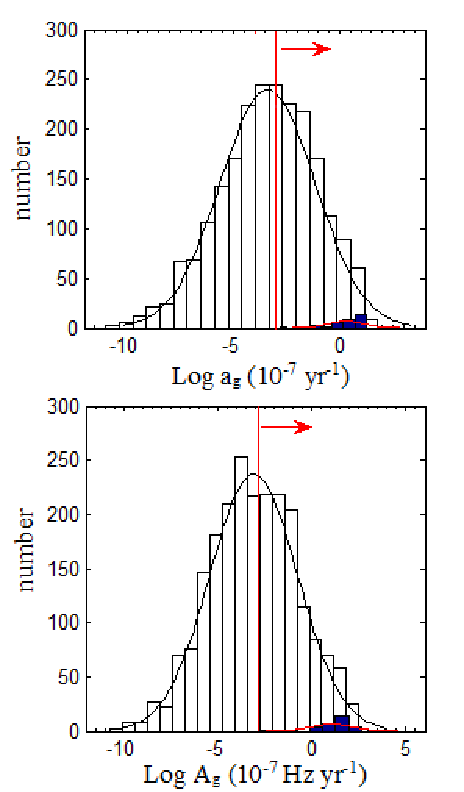}
    \caption{The distribution of glitch activity parameters in pulsars. The top panel is fractional glitch activity, while the bottom panel is absolute glitch activity. Shaded/coloured histograms are pulsars with multiple glitches, while plain histograms are for all the pulsars. The red vertical line indicate the  least observed activity parameter.}
    \label{fig:distribution_activity}
\end{figure}

To visualize how the activity parameters are distributed in the ensemble of pulsars, we examine the distribution of the activity parameters. 
The distribution plots are shown in Figure (\ref{fig:distribution_activity}). 
The distributions approximate a normal distribution as demonstrated by the fits on the plots. 
This is an indication that the equation used in the estimation describes the activity parameters well. 
The activity parameter of pulsars with multiple glitches is just a subset of the distribution of activity parameters of the ensemble of pulsars as expected (see the shaded histograms). 
More so, it shows that the activity parameters we measure today from pulsars of multiple glitches correspond only to the larger end of the activity parameter distribution.
 The red vertical lines with arrows indicate the least observed activity parameter. 
In the top panel, it corresponds to a cumulative fractional glitch size of,  $\Sigma \Delta\nu/\nu \sim 10^{-11} $ per  year and the bottom panel corresponds to a cumulative absolute glitch size of $\Sigma \Delta\nu \sim 1.0\times 10^{-8} $ Hz per year. 
Inevitably glitches of these sizes are difficult to measure because of their indistinguishable nature from timing noises.
So pulsars of very low spin-down rate ($|\dot{\nu}| \leq 10^{-14}$) could be glitching with glitches of such a sizes. 
When measurements of such small glitch sizes become feasible, the activity parameters of such pulsars shall be at the back end of the arrows.
This is also in line with the note of \cite{b7} who opined that the distribution of glitch sizes is incomplete at the small end of the glitch size distributions. 
In the future when glitch size distribution is complete, the distribution of the activity parameters will resemble that presented in Figure \ref{fig:distribution_activity}.

\begin{figure*}
\centering
\includegraphics[scale=0.8]{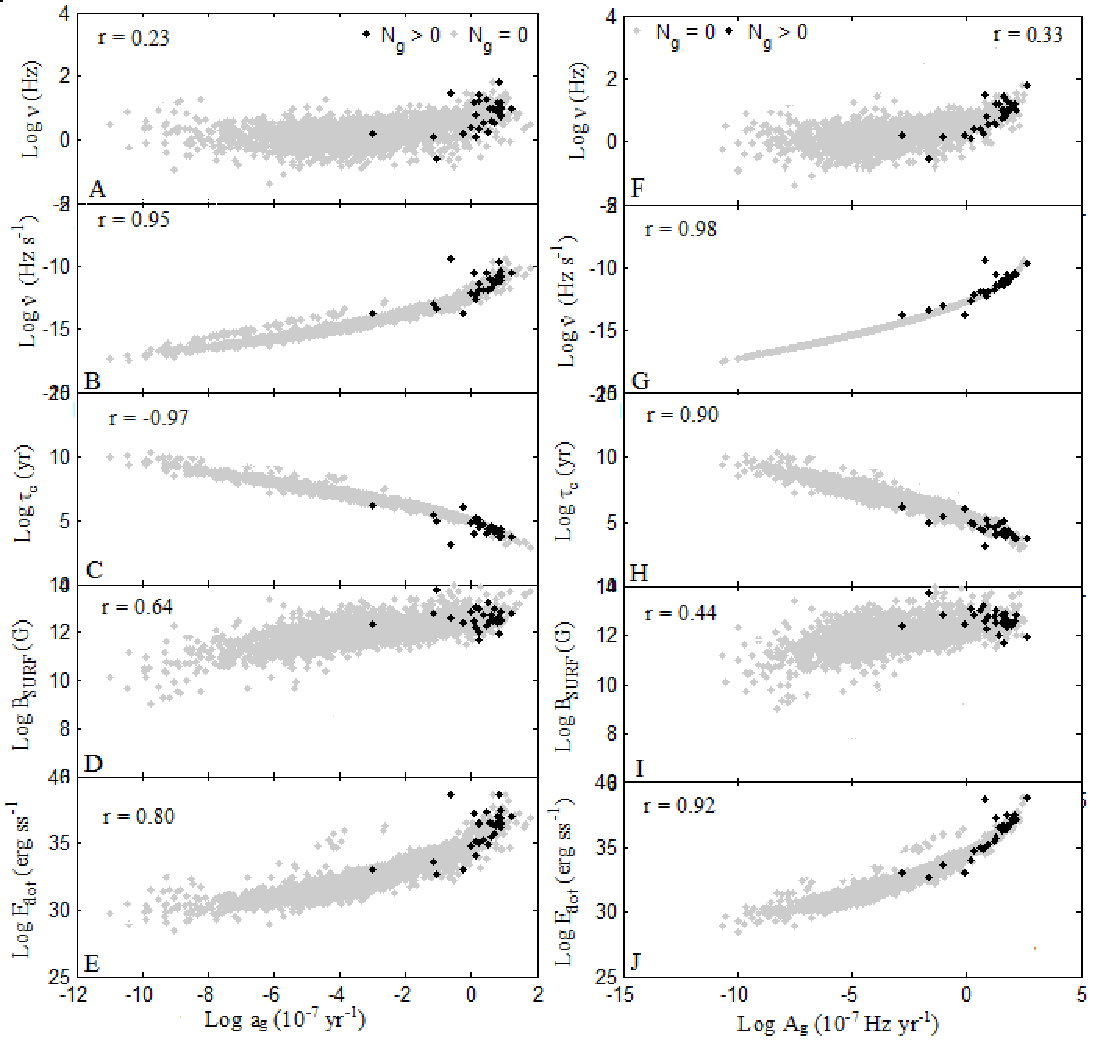}
    \caption{Activity parameter as a function of pulsar spin parameters. Panels: A and F are for spin frequency, B and G are for frequency derivatives, C and H are for characteristic age, D and I are for the surface magnetic field, and E and J are for spin-down luminosity. Black dots are for pulsars with multiple glitches, while grey dots are for all the pulsars. r denotes the correlation coefficient.}
    \label{fig:activity_vs_all.}
\end{figure*}

In what follows, we test the relationship between the estimated activity parameters with pulsar main parameters using the observed activity parameter as a control by plotting them together. 
The relationship is shown in Figure (\ref{fig:activity_vs_all.}).
In all the plots, the observed activity parameters fit well in the trend defined by the estimated activity parameters. 
This is a show of agreement between theoretical and observed activity parameters.
The spin frequency has a weak correlation with the activity parameters (as seen in panels A and F).  
 In this, we point out that in ensemble of pulsars, $ \nu $ does not have a strong relationship with activity parameters.
With the spin frequency derivative (panels B and G) the relationship with the activity parameters is very strong. 
The glitch activity increases with an increase in frequency derivative just as earlier reported by some authors \citep[e.g.][]{urama99,lyne00,Fuentes2017,eya22a}.
This finding is just another support to the notion that glitches are driven by pulsar spin-down rate.
In panels C and H, we can observe a strong inverse relationship between the activity parameters and the pulsar characteristic age, which is expected. 
This is because the characteristic age is partly determined by the spin frequency and its derivative, where the derivative is the denominator. 
As the activity parameters have a strong positive correlation with the frequency derivative, it can be seen that the effect of spin frequency on the activity parameter is minimal in the presence of a spin frequency derivative. 
This supports the idea that glitches are driven by spin-down rate, as older pulsars spin down less fast. 
Additionally, this inverse correlation is consistent with Figure (\ref{fig:age_size_figure}), which shows that glitch size and number of events decrease with pulsar characteristic age.
Panels D and I illustrate the relationship between the pulsar's surface magnetic field ($ \sim \sqrt{P\dot{P}} $) and the activity parameter. 
The correlation coefficient magnitude indicates that they exhibit a moderate relationship, which contradicts the report of \cite{Fuentes2017}. 
To obtain a clear picture of the correlation, observing more glitches in pulsars without records of glitches is necessary. 
Panels E and J show a strong correlation between the spin-down luminosity and the activity parameters. 
This is a pointer that the rate at which a pulsar loses rotational kinetic energy ($\sim I\dot{\nu}\nu$) influences the activity parameter. 
Indeed, parameters associated with frequency derivatives have a strong correlation with activity parameters, which highlights the importance of frequency derivatives in pulsar glitch events. The spin frequency derivative indicates the slowing down of a pulsar, and it is defined as $ \dot{\nu} = - K\nu^{n} $, where $n$ is the breaking index, and $K$ is a positive constant that depends on the pulsar's magnetic field strength, moment of inertia ($ I $), and the angle of inclination ($ \alpha $) between the rotation and magnetic axis. 
This equation is the standard spin-down law of pulsars. Assuming that pulsars spin similarly to a dipole rotator in a vacuum, $n$ is typically equal to 3. 
By integrating it and assuming that the pulsar's spin frequency at birth is far greater than what it is at present, we can determine the pulsar spin-down time, known as the characteristic age ($ \sim - \nu/2\dot{\nu} $). 
Therefore, we can see that the spin frequency derivatives are crucial components in the evolution of pulsars.

Finally, we have demonstrated that the glitch activity trend in pulsars without records of glitches is consistent with the trend defined by pulsars with records of multiple glitches. Therefore, the estimated glitch activity aligns well with the observation.

\section{Conclusions}
Pulsar glitches are useful tools in astrophysics as they help us explore the interior of neutron stars and gain insight into the behaviour of matter beyond nuclear densities. Several studies have been conducted in that direction [\citep*[e.g.][]{b53,Chamel12,Chamel13,Wlazlowski,eya17a,Hujeirat2019,Hujeirat2020}]. To further our understanding of the events, it is necessary to study the interplay between the interior dynamics of neutron stars and that of the observable crust. One way to do this is by examining the relationship between activity parameters and spin parameters. This analysis is a step toward that. Glitch sizes, which are an important aspect of activity parameters, depend on the interior dynamics of the neutron star, including the number of superfluid vortices involved, their migration distance after unpinning and repinning, and their location. Conversely, spin parameters describe the dynamics of the crust. Therefore, estimating the activity parameter from the spin parameter can provide a connection between the crust and the interior. 
Previous studies have concentrated on the linear relationship between the activity parameter and a selected spin parameter [\citep[e.g][]{b53,urama99,lyne00,wang00, Andersson2012,Fuentes2017,eya2019,eya22a}]. 
However, our analysis explores a different relationship between these parameters and finds that a quadratic fit provides a better fit to the data [see Figure (\ref{fig:ag-spindown_figure}) and Table (\ref{tab:fit_parameters}) especially the magnitude $ R^2 $ and $ n_{res} $]. This conclusion is in line with the fact that glitch activity peaks in middle-aged ($10^4 -10^5 yr$) pulsars \citep{urama99,b7}, a phenomenon that naturally fits a quadratic relationship (i.e. rising from pulsar birth, reaching maximum at middle age and declining after). 

In conclusion, our study demonstrates that it is possible to estimate the glitch activities of pulsars using their spin parameters, with the spin frequency derivative being the most reliable tool for this purpose. 
This approach can be helpful for theorists who need the glitch activity parameters of pulsars as input parameters while studying the interior of neutron stars.









\appendix
\begin{table*}
Table A: {Glitch activity parameters of pulsars. $ a_g$ is in unit of $\rm{yr^{-1}} $ and $A_g $ is in unit of Hz $ \rm{yr^{-1}}$}.

\begin{tiny}
\label{Tab:appendix}

\centering
\begin{tabular}{lccccccccccccccc} 
		\hline
	Pulsar	&$ a_{g} $	&$ A_{g} $	&	Pulsar	&		$ a_{g}$	&$ A_{g} $	& Pulsar	&	$ a_{g} $	&$ A_{g} $&Pulsar&$ a_{g} $&$ A_{g} $\\
\hline
J0002+6216	&	3.05E-08	&	2.64E-07	&	J0502+4654	&	7.79E-10	&	1.22E-09	&	J0821-4221	&	1.14E-09	&	2.87E-09	&	J1038-5831	&	3.79E-11	&	5.72E-11	\\
J0006+1834	&	9.43E-11	&	1.36E-10	&	J0502-6617	&	7.41E-09	&	1.07E-08	&	J0821-4300	&	3.24E-13	&	2.87E-12	&	J1041-1942	&	1.59E-12	&	1.15E-12	\\
J0007+7303	&	8.10E-07	&	2.56E-06	&	J0511-6508	&	8.60E-12	&	2.67E-11	&	J0826+2637	&	1.39E-10	&	2.62E-10	&	J1042-5521	&	2.01E-10	&	1.72E-10	\\
J0012+5431	&	2.74E-16	&	9.06E-17	&	J0514-4408	&	7.65E-10	&	2.39E-09	&	J0828-3417	&	6.02E-13	&	3.26E-13	&	J1043-6116	&	1.34E-08	&	4.64E-08	\\
J0014+4746	&	7.06E-13	&	5.69E-13	&	J0517+2212	&	4.24E-14	&	1.91E-13	&	J0831-4406	&	3.54E-10	&	1.14E-09	&	J1044-5737	&	2.83E-07	&	2.04E-06	\\
J0021-0909	&	2.74E-13	&	1.18E-13	&	J0518+5125	&	1.64E-13	&	1.80E-13	&	J0834-4159	&	1.84E-08	&	1.52E-07	&	J1046+0304	&	2.72E-12	&	8.34E-12	\\
J0024-7204ab	&	1.02E-14	&	2.75E-12	&	J0518+5416	&	4.60E-10	&	1.35E-09	&	J0835-3707	&	3.15E-09	&	5.81E-09	&	J1046-5813	&	1.81E-10	&	4.91E-10	\\
\end{tabular}
\end{tiny}
\end{table*}

\begin{table*}

Table A continues: {Glitch activity parameters of pulsars.}
\begin{tiny}

\centering
	\begin{tabular}{lccccccccccccccc} 
		\hline
	Pulsar	&$ a_{g} $	&$ A_{g} $	&	Pulsar	&		$ a_{g}$	&$ A_{g} $	& Pulsar	&	$ a_{g} $	&$ A_{g} $&Pulsar&$ a_{g} $&$ A_{g} $\\
\hline
J0139+5621	&	6.35E-09	&	3.58E-09	&	J0614+2229	&	1.09E-07	&	3.26E-07	&	J0922+0638	&	9.07E-09	&	2.11E-08	&	J1107-5907	&	1.33E-14	&	5.26E-14	\\
J0139+5814	&	1.56E-08	&	5.72E-08	&	J0621+0336	&	5.72E-15	&	2.12E-14	&	J0922-4949	&	3.56E-08	&	3.74E-08	&	J1107-5947	&	9.36E-14	&	6.17E-14	\\
J0141+6009	&	3.05E-13	&	2.49E-13	&	J0622+3749	&	3.65E-08	&	1.09E-07	&	J0924-5302	&	1.24E-08	&	1.65E-08	&	J1107-6143	&	1.91E-08	&	1.06E-08	\\
J0147+5922	&	6.17E-11	&	3.14E-10	&	J0623+0340	&	1.17E-13	&	1.91E-13	&	J0924-5814	&	4.14E-10	&	5.60E-10	&	J1110-5637	&	1.73E-10	&	3.10E-10	\\
J0151-0635	&	2.03E-13	&	1.38E-13	&	J0624-0424	&	3.38E-12	&	3.25E-12	&	J0927+2345	&	1.02E-12	&	1.34E-12	&	J1112-6103	&	3.11E-07	&	4.78E-06	\\
J0152+0948	&	4.73E-13	&	1.72E-13	&	J0627+0649	&	4.48E-10	&	1.29E-09	&	J0930-2301	&	1.14E-11	&	6.31E-12	&	J1112-6613	&	1.28E-10	&	3.84E-10	\\
J0152-1637	&	1.95E-11	&	2.35E-11	&	J0627+0706	&	2.38E-08	&	5.01E-08	&	J0931-1902	&	3.86E-16	&	8.32E-14	&	J1112-6926	&	1.03E-10	&	1.26E-10	\\
\end{tabular}
\end{tiny}
\end{table*}

\begin{table*}

Table A continues: {Glitch activity parameters of pulsars.}
\begin{tiny}

\centering
	\begin{tabular}{lccccccccccccccc} 
		\hline
	Pulsar	&$ a_{g} $	&$ A_{g} $	&	Pulsar	&		$ a_{g}$	&$ A_{g} $	& Pulsar	&	$ a_{g} $	&$ A_{g} $&Pulsar&$ a_{g} $&$ A_{g} $\\
\hline
J0415+6954	&	4.70E-13	&	1.20E-12	&	J0754+3231	&	1.88E-12	&	1.30E-12	&	J1019-5749	&	8.17E-08	&	5.02E-07	&	J1148-5725	&	1.56E-11	&	4.38E-12	\\
J0418-4154	&	2.76E-11	&	3.64E-11	&	J0758-1528	&	5.86E-11	&	8.59E-11	&	J1020-5921	&	4.83E-09	&	3.90E-09	&	J1148-6415	&	7.97E-13	&	2.46E-13	\\
J0421+3255	&	8.75E-15	&	9.72E-15	&	J0803-0942	&	5.14E-11	&	9.00E-11	&	J1020-6026	&	2.62E-08	&	1.87E-07	&	J1151-6108	&	6.96E-08	&	6.85E-07	\\
J0421-0345	&	4.78E-13	&	2.21E-13	&	J0804-3647	&	9.59E-12	&	4.38E-12	&	J1021-5601	&	2.28E-14	&	3.40E-14	&	J1152-5800	&	1.26E-12	&	7.01E-13	\\
J0426+4933	&	9.20E-09	&	9.97E-09	&	J0807-5421	&	6.19E-12	&	1.18E-11	&	J1022-5813	&	2.09E-08	&	1.27E-08	&	J1152-6012	&	3.88E-09	&	1.03E-08	\\
J0435+2749	&	3.49E-15	&	1.07E-14	&	J0808-3937	&	1.21E-12	&	1.39E-12	&	J1023-5746	&	1.41E-06	&	1.27E-05	&	J1154-6250	&	9.84E-11	&	3.49E-10	\\
J0448-2749	&	1.30E-12	&	2.89E-12	&	J0809-4753	&	3.93E-10	&	7.18E-10	&	J1024-0719	&	1.33E-14	&	2.58E-12	&	J1156-5707	&	4.93E-08	&	1.71E-07	\\
\end{tabular}
\end{tiny}
\end{table*}

\begin{table*}

Table A continues: {Glitch activity parameters of pulsars.}
\begin{tiny}

\centering
	\begin{tabular}{lccccccccccccccc} 
		\hline
	Pulsar	&$ a_{g} $	&$ A_{g} $	&	Pulsar	&		$ a_{g}$	&$ A_{g} $	& Pulsar	&	$ a_{g} $	&$ A_{g} $&Pulsar&$ a_{g} $&$ A_{g} $\\
\hline
J1222-5738	&	4.62E-14	&	4.27E-14	&	J1350-5115	&	1.54E-10	&	5.20E-10	&	J1514-5316	&	5.60E-17	&	1.89E-16	&	J1600-5916	&	2.03E-12	&	1.62E-12	\\
J1223-5856	&	2.24E-15	&	7.75E-15	&	J1350-6225	&	3.83E-08	&	2.77E-07	&	J1514-5925	&	7.32E-09	&	4.92E-08	&	J1601-5244	&	7.09E-14	&	2.77E-14	\\
J1224-6208	&	8.61E-09	&	1.47E-08	&	J1352-6803	&	4.32E-11	&	6.87E-11	&	J1515-5720	&	6.06E-09	&	2.11E-08	&	J1601-5335	&	1.42E-07	&	4.94E-07	\\
J1224-6407	&	7.84E-09	&	3.62E-08	&	J1354-6249	&	5.62E-11	&	1.90E-11	&	J1517-4356	&	8.09E-13	&	1.24E-12	&	J1602-4957	&	2.66E-09	&	3.24E-09	\\
J1225-5556	&	2.72E-11	&	2.67E-11	&	J1355-5153	&	2.09E-10	&	3.25E-10	&	J1517-4636	&	4.41E-11	&	4.97E-11	&	J1602-5100	&	2.58E-08	&	2.99E-08	\\
J1225-6035	&	1.81E-12	&	2.88E-12	&	J1355-5747	&	1.78E-13	&	8.74E-14	&	J1518+0204A	&	6.25E-14	&	1.13E-11	&	J1603-2531	&	6.75E-10	&	2.39E-09	\\
J1225-6408	&	8.64E-11	&	2.06E-10	&	J1355-5925	&	1.43E-10	&	1.18E-10	&	J1518-0627	&	1.82E-12	&	2.29E-12	&	J1603-2712	&	1.37E-10	&	1.77E-10	\\
\end{tabular}
\end{tiny}
\end{table*}

\begin{table*}

Table A continues: {Glitch activity parameters of pulsars.}
\begin{tiny}

\centering
	\begin{tabular}{lccccccccccccccc} 
		\hline
	Pulsar	&$ a_{g} $	&$ A_{g} $	&	Pulsar	&		$ a_{g}$	&$ A_{g} $	& Pulsar	&	$ a_{g} $	&$ A_{g} $&Pulsar&$ a_{g} $&$ A_{g} $\\
\hline
J1316-6232	&	3.28E-09	&	9.57E-09	&	J1435-5954	&	1.60E-10	&	3.38E-10	&	J1541+47	&	1.44E-11	&	5.17E-11	&	J1622-4845	&	9.83E-14	&	1.34E-13	\\
J1317-5759	&	5.68E-11	&	2.15E-11	&	J1437-5959	&	1.01E-07	&	1.63E-06	&	J1541-5535	&	1.71E-07	&	5.78E-07	&	J1622-4944	&	1.52E-09	&	1.41E-09	\\
J1317-6302	&	3.79E-12	&	1.45E-11	&	J1437-6146	&	2.14E-09	&	4.57E-09	&	J1542-5034	&	5.06E-10	&	8.44E-10	&	J1623-0841	&	2.11E-10	&	4.19E-10	\\
J1319-6056	&	6.19E-10	&	2.18E-09	&	J1440-6344	&	9.24E-11	&	2.01E-10	&	J1542-5133	&	1.88E-13	&	1.05E-13	&	J1623-0908	&	2.08E-11	&	1.63E-11	\\
J1319-6105	&	2.10E-10	&	4.98E-10	&	J1441-6137	&	2.89E-13	&	2.46E-13	&	J1542-5303	&	1.52E-08	&	1.26E-08	&	J1623-4256	&	1.46E-10	&	4.01E-10	\\
J1320-3512	&	1.11E-17	&	2.43E-17	&	J1443-5122	&	1.51E-12	&	2.06E-12	&	J1543+0929	&	2.45E-12	&	3.28E-12	&	J1623-4949	&	1.72E-08	&	2.37E-08	\\
J1320-5359	&	1.20E-08	&	4.27E-08	&	J1444-5941	&	1.94E-11	&	7.01E-12	&	J1543-0620	&	1.44E-11	&	2.03E-11	&	J1624+5850	&	2.18E-12	&	3.35E-12	\\
\end{tabular}
\end{tiny}
\end{table*}

\begin{table*}

Table A continues: {Glitch activity parameters of pulsars.}
\begin{tiny}

\centering
	\begin{tabular}{lccccccccccccccc} 
		\hline
	Pulsar	&$ a_{g} $	&$ A_{g} $	&	Pulsar	&		$ a_{g}$	&$ A_{g} $	& Pulsar	&	$ a_{g} $	&$ A_{g} $&Pulsar&$ a_{g} $&$ A_{g} $\\
\hline
J1638-3815	&	4.92E-14	&	7.05E-14	&	J1705-3423	&	4.36E-10	&	1.71E-09	&	J1733-2228	&	4.04E-15	&	4.63E-15	&	J1750-3503	&	8.57E-15	&	1.25E-14	\\
J1638-3951	&	4.45E-12	&	5.77E-12	&	J1705-3936	&	3.34E-09	&	3.90E-09	&	J1733-2533	&	1.10E-10	&	1.66E-10	&	J1750-3703D	&	1.09E-11	&	2.13E-09	\\
J1638-4233	&	3.80E-09	&	7.43E-09	&	J1705-3950	&	1.20E-07	&	3.76E-07	&	J1733-2837	&	1.81E-11	&	2.36E-11	&	J1751-2516	&	7.10E-10	&	1.80E-09	\\
J1638-4344	&	2.87E-16	&	2.56E-16	&	J1705-4108	&	8.76E-09	&	1.02E-08	&	J1733-3030	&	3.78E-10	&	1.04E-09	&	J1751-3323	&	2.57E-09	&	4.69E-09	\\
J1638-4417	&	4.88E-09	&	4.14E-08	&	J1705-4331	&	2.94E-12	&	1.32E-11	&	J1733-3322	&	6.07E-11	&	4.87E-11	&	J1751-4657	&	2.84E-11	&	3.82E-11	\\
J1638-4608	&	1.19E-07	&	4.29E-07	&	J1705-6135	&	1.40E-14	&	1.73E-14	&	J1733-3716	&	1.70E-08	&	5.03E-08	&	J1752+2359	&	4.30E-11	&	1.05E-10	\\
J1638-5226	&	1.04E-09	&	3.04E-09	&	J1706-3839	&	3.10E-10	&	5.28E-10	&	J1733-4005	&	4.96E-10	&	8.83E-10	&	J1752-2410	&	3.42E-10	&	1.79E-09	\\
\end{tabular}
\end{tiny}
\end{table*}

\begin{table*}

Table A continues: {Glitch activity parameters of pulsars.}
\begin{tiny}

\centering
	\begin{tabular}{lccccccccccccccc} 
		\hline
	Pulsar	&$ a_{g} $	&$ A_{g} $	&	Pulsar	&		$ a_{g}$	&$ A_{g} $	& Pulsar	&	$ a_{g} $	&$ A_{g} $&Pulsar&$ a_{g} $&$ A_{g} $\\
\hline
J1646-4308	&	6.12E-14	&	7.27E-14	&	J1710-2616	&	3.22E-16	&	3.37E-16	&	J1738-2647	&	1.31E-09	&	3.75E-09	&	J1756-2435	&	1.39E-12	&	2.08E-12	\\
J1646-4346	&	3.53E-07	&	1.52E-06	&	J1710-4148	&	2.73E-12	&	9.54E-12	&	J1738-2736	&	5.77E-10	&	9.19E-10	&	J1756-2619	&	2.78E-11	&	3.83E-11	\\
J1646-5123	&	2.08E-10	&	3.92E-10	&	J1711-1509	&	1.19E-11	&	1.37E-11	&	J1738-2955	&	1.07E-07	&	2.41E-07	&	J1757-1500	&	1.82E-09	&	1.02E-08	\\
J1646-6831	&	2.48E-12	&	1.39E-12	&	J1711-3826	&	2.82E-09	&	6.07E-09	&	J1738-3107	&	3.10E-12	&	5.64E-12	&	J1757-2223	&	5.60E-10	&	3.02E-09	\\
J1647+6608	&	1.06E-10	&	6.62E-11	&	J1711-5350	&	2.02E-09	&	2.24E-09	&	J1738-3211	&	8.79E-12	&	1.14E-11	&	J1757-2421	&	2.69E-08	&	1.15E-07	\\
J1647-3607	&	1.22E-11	&	5.74E-11	&	J1712-2715	&	5.95E-10	&	2.33E-09	&	J1738-3316	&	5.84E-14	&	7.99E-14	&	J1758+3030	&	3.39E-12	&	3.58E-12	\\
J1648-3256	&	2.37E-10	&	3.29E-10	&	J1713+7810	&	8.23E-13	&	1.90E-12	&	J1739+0612	&	1.34E-11	&	5.71E-11	&	J1758-1931	&	4.42E-09	&	6.38E-09	\\
\end{tabular}
\end{tiny}
\end{table*}

\begin{table*}

Table A continues: {Glitch activity parameters of pulsars.}
\begin{tiny}

\centering
	\begin{tabular}{lccccccccccccccc} 
		\hline
	Pulsar	&$ a_{g} $	&$ A_{g} $	&	Pulsar	&		$ a_{g}$	&$ A_{g} $	& Pulsar	&	$ a_{g} $	&$ A_{g} $&Pulsar&$ a_{g} $&$ A_{g} $\\
\hline
J1657+3304	&	4.15E-12	&	2.65E-12	&	J1722-4400	&	9.37E-11	&	4.29E-10	&	J1745-3040	&	8.57E-09	&	2.33E-08	&	J1803-2137	&	6.44E-07	&	4.82E-06	\\
J1657-4432	&	1.75E-09	&	2.87E-09	&	J1723-2852	&	2.27E-11	&	3.63E-11	&	J1745-3812	&	1.24E-10	&	1.77E-10	&	J1803-2149	&	1.32E-07	&	1.24E-06	\\
J1658-4306	&	6.14E-09	&	5.26E-09	&	J1723-3659	&	1.78E-08	&	8.79E-08	&	J1746+2245	&	2.57E-12	&	7.43E-13	&	J1803-3002A	&	2.48E-14	&	3.49E-12	\\
J1658-4958	&	1.21E-09	&	2.89E-09	&	J1724-3149	&	4.30E-10	&	4.54E-10	&	J1746+2540	&	5.37E-12	&	5.07E-12	&	J1803-3329	&	2.52E-12	&	3.98E-12	\\
J1804-2228	&	5.03E-13	&	8.81E-13	&	J1821-1419	&	2.91E-07	&	1.76E-07	&	J1835-1106	&	8.18E-08	&	4.93E-07	&	J1846-0749	&	2.52E-10	&	7.20E-10	\\
J1805+0306	&	5.69E-10	&	2.60E-09	&	J1821-1432	&	2.64E-11	&	1.38E-11	&	J1835-1548	&	7.16E-11	&	1.07E-10	&	J1846-07492	&	2.97E-10	&	3.45E-10	\\
J1805-0619	&	7.15E-11	&	1.57E-10	&	J1822+0705	&	7.01E-12	&	5.14E-12	&	J1836-0436	&	4.06E-10	&	1.15E-09	&	J1846-7403	&	1.14E-12	&	2.34E-13	\\
\end{tabular}
\end{tiny}
\end{table*}

\begin{table*}

Table A continues: {Glitch activity parameters of pulsars.}
\begin{tiny}

\centering
	\begin{tabular}{lccccccccccccccc} 
		\hline
	Pulsar	&$ a_{g} $	&$ A_{g} $	&	Pulsar	&		$ a_{g}$	&$ A_{g} $	& Pulsar	&	$ a_{g} $	&$ A_{g} $&Pulsar&$ a_{g} $&$ A_{g} $\\
\hline
J1811-2439	&	8.11E-12	&	1.95E-11	&	J1826-2415	&	1.61E-14	&	3.43E-12	&	J1840-1122	&	3.47E-10	&	3.69E-10	&	J1851+1259	&	5.30E-10	&	4.40E-10	\\
J1811-4930	&	1.04E-11	&	7.28E-12	&	J1827-0750	&	7.17E-10	&	2.65E-09	&	J1840-1207	&	1.70E-10	&	2.25E-10	&	J1851-0029	&	9.97E-10	&	1.92E-09	\\
J1812+0226	&	1.85E-10	&	2.33E-10	&	J1827-0934	&	2.15E-09	&	4.19E-09	&	J1840-1419	&	3.80E-13	&	5.75E-14	&	J1851-0053	&	1.24E-12	&	8.81E-13	\\
J1812-1718	&	1.36E-09	&	1.13E-09	&	J1827-0958	&	4.25E-10	&	1.73E-09	&	J1841+0912	&	1.51E-10	&	3.95E-10	&	J1851-0114	&	4.98E-11	&	5.22E-11	\\
J1812-1733	&	4.39E-11	&	8.15E-11	&	J1827-1446	&	3.97E-08	&	7.96E-08	&	J1841-0157	&	5.45E-09	&	8.22E-09	&	J1851-0241	&	3.72E-09	&	8.54E-09	\\
J1812-1910	&	4.00E-08	&	9.28E-08	&	J1828+1221	&	4.65E-12	&	3.04E-12	&	J1841-0310	&	6.01E-14	&	3.62E-14	&	J1851-0633	&	4.24E-12	&	2.21E-12	\\
J1812-20	&	1.38E-14	&	7.23E-15	&	J1828+1359	&	8.19E-12	&	1.10E-11	&	J1841-0345	&	2.02E-07	&	9.89E-07	&	J1852+0008	&	1.77E-09	&	3.79E-09	\\
\end{tabular}
\end{tiny}
\end{table*}

\begin{table*}

Table A continues: {Glitch activity parameters of pulsars.}
\begin{tiny}

\centering
	\begin{tabular}{lccccccccccccccc} 
		\hline
	Pulsar	&$ a_{g} $	&$ A_{g} $	&	Pulsar	&		$ a_{g}$	&$ A_{g} $	& Pulsar	&	$ a_{g} $	&$ A_{g} $&Pulsar&$ a_{g} $&$ A_{g} $\\
\hline
J1818-1541	&	2.96E-09	&	5.38E-09	&	J1833-6023	&	1.97E-14	&	1.04E-14	&	J1844-0030	&	9.00E-10	&	1.40E-09	&	J1855-0941	&	9.43E-12	&	2.73E-11	\\
J1818-1556	&	3.18E-12	&	3.34E-12	&	J1834-0031	&	4.02E-11	&	1.22E-10	&	J1844-0244	&	8.72E-09	&	1.72E-08	&	J1856+0102	&	4.42E-11	&	7.13E-11	\\
J1819+1305	&	4.32E-13	&	4.07E-13	&	J1834-0426	&	1.20E-12	&	4.14E-12	&	J1844-0302	&	2.53E-10	&	2.11E-10	&	J1856+0113	&	5.61E-07	&	2.10E-06	\\
J1819-0925	&	1.13E-10	&	1.33E-10	&	J1834-0602	&	2.02E-10	&	4.14E-10	&	J1844-0310	&	3.65E-09	&	6.95E-09	&	J1856+0245	&	4.67E-07	&	5.78E-06	\\
J1819-1008	&	4.10E-10	&	1.36E-09	&	J1834-0633	&	8.13E-11	&	2.56E-10	&	J1844-0346	&	7.83E-07	&	6.94E-06	&	J1856+0404	&	6.08E-14	&	1.45E-13	\\
J1819-1114	&	8.92E-11	&	3.03E-10	&	J1834-0731	&	5.36E-08	&	1.04E-07	&	J1844-0433	&	1.12E-10	&	1.13E-10	&	J1856+0912	&	3.35E-12	&	1.54E-12	\\
J1819-1131	&	9.45E-13	&	6.81E-13	&	J1834-0742	&	9.58E-09	&	1.22E-08	&	J1844-0452	&	1.63E-10	&	6.04E-10	&	J1856-0526	&	3.75E-10	&	1.01E-09	\\
\end{tabular}
\end{tiny}
\end{table*}

\begin{table*}

Table A continues: {Glitch activity parameters of pulsars.}
\begin{tiny}

\centering
	\begin{tabular}{lccccccccccccccc} 
		\hline
	Pulsar	&$ a_{g} $	&$ A_{g} $	&	Pulsar	&		$ a_{g}$	&$ A_{g} $	& Pulsar	&	$ a_{g} $	&$ A_{g} $&Pulsar&$ a_{g} $&$ A_{g} $\\
\hline

J1901+0459	&	2.19E-09	&	2.49E-09	&	J1911+0921	&	5.18E-14	&	1.89E-13	&	J1927+1856	&	1.08E-09	&	3.61E-09	&	J1955+5059	&	9.64E-11	&	1.86E-10	\\
J1901+0510	&	1.52E-08	&	2.48E-08	&	J1911+0925	&	1.93E-09	&	5.95E-09	&	J1927+2234	&	8.68E-13	&	6.06E-13	&	J1955+6708	&	7.46E-16	&	8.70E-14	\\
J1901+0511	&	4.06E-11	&	8.83E-12	&	J1911+1051	&	3.46E-08	&	1.81E-07	&	J1928+1443	&	1.39E-13	&	1.38E-13	&	J1956+0838	&	1.19E-11	&	3.91E-11	\\
J1901+0621	&	4.52E-16	&	5.43E-16	&	J1911+1301	&	2.31E-11	&	2.29E-11	&	J1928+1746	&	1.38E-07	&	2.01E-06	&	J1957+2831	&	1.74E-09	&	5.66E-09	\\
J1901+0716	&	1.40E-10	&	2.17E-10	&	J1911+1336	&	6.17E-12	&	2.06E-11	&	J1928+1923	&	5.07E-10	&	6.20E-10	&	J1957+5033	&	4.30E-09	&	1.15E-08	\\
J1901+1306	&	3.11E-15	&	1.70E-15	&	J1911+1347	&	1.60E-14	&	3.47E-12	&	J1928-0108	&	2.82E-12	&	1.19E-12	&	J1957-0002	&	4.10E-12	&	4.24E-12	\\
J1901-0125	&	4.21E-13	&	1.51E-10	&	J1911+1758	&	2.78E-15	&	6.04E-15	&	J1929+1357	&	1.47E-10	&	1.70E-10	&	J1958+2846	&	5.28E-07	&	1.82E-06	\\
\end{tabular}
\end{tiny}
\end{table*}

\begin{table*}

Table A continues: {Glitch activity parameters of pulsars.}
\begin{tiny}

\centering
	\begin{tabular}{lccccccccccccccc} 
		\hline
	Pulsar	&$ a_{g} $	&$ A_{g} $	&	Pulsar	&		$ a_{g}$	&$ A_{g} $	& Pulsar	&	$ a_{g} $	&$ A_{g} $&Pulsar&$ a_{g} $&$ A_{g} $\\
\hline
J1905+0400	&	1.83E-15	&	4.82E-13	&	J1915+1009	&	1.21E-08	&	3.00E-08	&	J1936+2042	&	5.13E-09	&	3.69E-09	&	J2013+3058	&	7.41E-12	&	2.68E-11	\\
J1905+0600	&	1.03E-10	&	2.33E-10	&	J1915+1145	&	2.15E-13	&	1.24E-12	&	J1937+1505	&	7.03E-12	&	2.45E-12	&	J2013+3845	&	1.62E-08	&	7.05E-08	\\
J1905+0616	&	5.31E-08	&	5.37E-08	&	J1915+1150	&	9.72E-08	&	9.71E-07	&	J1937+2544	&	3.22E-10	&	1.60E-09	&	J2013-0649	&	1.23E-11	&	2.12E-11	\\
J1905+0709	&	5.99E-10	&	9.24E-10	&	J1915+1410	&	4.50E-13	&	1.51E-12	&	J1937+2950	&	1.66E-11	&	1.00E-11	&	J2015+2524	&	4.46E-14	&	1.94E-14	\\
J1905+0902	&	4.54E-09	&	2.08E-08	&	J1915+1647	&	1.09E-13	&	6.73E-14	&	J1938+0650	&	1.64E-10	&	1.46E-10	&	J2017+2043	&	4.54E-11	&	8.46E-11	\\
J1905+1034	&	5.88E-10	&	3.40E-10	&	J1916+0748	&	3.66E-09	&	6.75E-09	&	J1938+2010	&	2.51E-10	&	3.65E-10	&	J2017+3625	&	1.82E-09	&	1.09E-08	\\
J1905-0056	&	2.47E-10	&	3.85E-10	&	J1916+0844	&	6.33E-10	&	1.44E-09	&	J1938+2213	&	1.83E-07	&	1.10E-06	&	J2017+5906	&	4.73E-12	&	1.17E-11	\\
\end{tabular}
\end{tiny}
\end{table*}

\begin{table*}

Table A continues: {Glitch activity parameters of pulsars.}
\begin{tiny}

	\begin{tabular}{lccccc} 
		\hline
	Pulsar	&$ a_{g} $	&$ A_{g} $	&Pulsar	&$ a_{g} $	&$ A_{g} $\\
\hline
J2053+4718	&	9.49E-12	&	1.93E-12	&		J2212+2933	&	1.15E-12	&	1.14E-12	\\
J2053-7200	&	6.47E-12	&	1.89E-11	&		J2215+1538	&	6.69E-10	&	1.79E-09	\\
J2055+15	&	2.89E-13	&	1.34E-10	&		J2215+4524	&	8.36E-12	&	3.07E-12	\\
J2055+2209	&	2.24E-11	&	2.75E-11	&		J2216+5759	&	9.38E-08	&	2.24E-07	\\
J2055+2539	&	2.51E-09	&	7.86E-09	&		J2217+5733	&	1.78E-12	&	1.68E-12	\\
J2055+3630	&	8.80E-11	&	3.97E-10	&		J2219+4754	&	3.36E-10	&	6.24E-10	\\
J2057+4701	&	4.35E-08	&	7.78E-08	&		J2222+2923	&	3.14E-15	&	1.11E-14	\\
\end{tabular}
\end{tiny}
\end{table*}
\label{lastpage}

\end{document}